\documentclass[fleqn,usenatbib]{mnras}

\usepackage{newtxtext,newtxmath}

\usepackage[T1]{fontenc}

\DeclareRobustCommand{\VAN}[3]{#2}
\let\VANthebibliography\thebibliography
\def\thebibliography{\DeclareRobustCommand{\VAN}[3]{##3}\VANthebibliography}

\usepackage{graphicx}	
\usepackage{amsmath}	
\usepackage{xspace}
\usepackage{ulem}
\usepackage{threeparttable}
\usepackage{booktabs}
\usepackage{orcidlink}

\newcommand{\alloa}[3]{\hbox{\ion{#1}{#2}$\lambda$#3}\xspace}
\newcommand{\forba}[3]{\hbox{[\ion{#1}{#2}]$\lambda$#3}\xspace}

\newcommand{\rforba}[4]{\hbox{[\ion{#1}{#2}]$\lambda$#3/$\lambda$#4}\xspace}

\newcommand{\Ha}{\hbox{H$\alpha$}\xspace}
\newcommand{\Hb}{\hbox{H$\beta$}\xspace}
\newcommand{\LHa}{\hbox{LH$\alpha$}\xspace}
\newcommand{\LHagal}{\hbox{LH$\alpha_{\mathrm{gal}}$}\xspace}
\newcommand{\aLF}{\hbox{$\alpha_\mathrm{LF}$}\xspace}
\newcommand{\hii}{\hbox{\ion{H}{ii}}\xspace}
\newcommand{\Nhii}{\hbox{N$_\ion{H}{ii}$}\xspace}

\newcommand{\Nregmin}{\hbox{$N_\mathrm{HII,min}$}\xspace}

\newcommand{\ksid}{\hbox{$\xi_{\rm{d}}$}\xspace}
\newcommand{\ksidsol}{\hbox{$\xi_{\rm{d},\odot}$}\xspace}
\newcommand{\Zsun}{\hbox{${Z}_{\odot}$}\xspace}
\newcommand{\HbFrac}{\hbox{H$\beta$Frac}\xspace}
\newcommand{\DIGFrac}{\hbox{DIGFrac}\xspace}
\newcommand{\cloudy}{\texttt{CLOUDY}\xspace}
\newcommand{\Cue}{\texttt{Cue}\xspace}
\newcommand{\pycloudy}{\texttt{pyCloudy}\xspace}
\newcommand{\pyneb}{\texttt{PyNeb}\xspace}

\newcommand{\logOH}{\hbox{$12 + \log\mathrm{(O/H)}$}\xspace}
\newcommand{\logOHgal}{\hbox{$12 + \log\mathrm{(O/H)_{gal}}$}\xspace}
\newcommand{\OHgal}{\hbox{$\mathrm{(O/H)_{gal}}$}\xspace}
\newcommand{\OH}{\hbox{$\mathrm{O/H}$}\xspace}
\newcommand{\OHtot}{\hbox{$\mathrm{(O/H)^{tot}}$}\xspace}

\newcommand{\NO}{\hbox{$\mathrm{N/O}$}\xspace}
\newcommand{\NHtot}{\hbox{$\mathrm{(N/H)^{tot}}$}\xspace}
\newcommand{\NOtot}{\hbox{$\mathrm{(N/O)^{tot}}$}\xspace}
\newcommand{\CO}{\hbox{$\mathrm{C/O}$}\xspace}
\newcommand{\NeO}{\hbox{$\mathrm{Ne/O}$}\xspace}
\newcommand{\SO}{\hbox{$\mathrm{S/O}$}\xspace}
\newcommand{\ArO}{\hbox{$\mathrm{Ar/O}$}\xspace}
\newcommand{\XO}{\hbox{$\mathrm{X/O}$}\xspace}
\newcommand{\XOgal}{\hbox{$\mathrm{(X/O)_{gal}}$}\xspace}
\newcommand{\CHtot}{\hbox{$\mathrm{(C/H)^{tot}}$}\xspace}
\newcommand{\COtot}{\hbox{$\mathrm{(C/O)^{tot}}$}\xspace}
\newcommand{\SHtot}{\hbox{$\mathrm{(S/H)^{tot}}$}\xspace}
\newcommand{\SOtot}{\hbox{$\mathrm{(S/O)^{tot}}$}\xspace}

\newcommand{\NeOtot}{\hbox{$\mathrm{(Ne/O)^{tot}}$}\xspace}

\newcommand{\ArOtot}{\hbox{$\mathrm{(Ar/O)^{tot}}$}\xspace}
\newcommand{\XHtot}{\hbox{$\mathrm{(X/H)^{tot}}$}\xspace}
\newcommand{\logU}{\hbox{$\log U$}\xspace}
\newcommand{\nh}{\hbox{$n_{\mathrm{H}}$}\xspace}

\newcommand{\fr}{\hbox{$f_\mathrm{r}$}\xspace}
\newcommand{\fc}{\hbox{$f_\mathrm{c}$}\xspace}
\newcommand{\fesc}{\hbox{$f_\mathrm{esc}$}\xspace}
\newcommand{\Te}{\hbox{$T_\mathrm{e}$}\xspace}

\newcommand{\NHIIBPB}{2,001,534\xspace}

\newcommand{\NHIItrain}{800,000\xspace}

\newcommand{\Nsyngals}{250,000\xspace}
\newcommand{\NbrightOiii}{207,750\xspace}
\newcommand{\NSDSS}{28,311\xspace}

\title[Nebular emission from composite galaxies]{Nebular emission from composite star-forming galaxies -- I. A novel modelling approach}
\author[C. Morisset et al.]{Christophe Morisset$^{\orcidlink{0000-0001-5801-6724}}$,$^{1, 2}$\thanks{E-mail: chris.morisset@gmail.com}
Stéphane Charlot$^{\orcidlink{0000-0003-3458-2275}}$,$^{3}$ Sebastián  F. Sánchez$^{\orcidlink{0000-0001-6444-9307}}$,$^{1}$ 
Carlos Espinosa-Ponce$^{\orcidlink{0000-0002-9658-8886}}$,$^{4}$ 
\newauthor
Eric Barat$^{\orcidlink{0000-0003-1113-5245}}$ $^{5}$ and Thomas Dautremer$^{\orcidlink{0000-0002-5035-4523}}$ $^{5}$\\
$^{1}$Universidad Nacional Autónoma de México, Instituto de Astronomía, AP 106,  Ensenada 22800, BC, México\\
$^2$Instituto de Ciencias Físicas, Universidad Nacional Autónoma de México, Av. Universidad s/n, 62210 Cuernavaca, Mor., México\\
$^{3}$Sorbonne Universit\'e, CNRS, UMR7095, Institut d'Astrophysique de Paris, F-75014, Paris, France\\
$^4$Instituto de Astronom\'ia, Universidad Nacional Aut\'onoma de  M\'exico, A.~P. 70-264, C.P. 04510, México, D.F., Mexico.\\
$^5$ Université Paris-Saclay, CEA, List, F-91120 Palaiseau, France
}

\date{Accepted XXX. Received YYY; in original form ZZZ}

\pubyear{2022}

\begin{document}
\label{firstpage}
\pagerange{\pageref{firstpage}--\pageref{lastpage}}
\maketitle

\begin{abstract}
We introduce a novel approach to modelling the nebular emission from star-forming galaxies by combining the contributions from many \hii regions incorporating loose trends in physical properties, random dust attenuation, a predefined \Ha luminosity function and a diffuse ionized-gas component. Using a machine-learning-based regression artificial neural network trained on a grid of models generated by the photoionization code \cloudy, we efficiently predict emission-line properties of individual \hii regions over a wide range of physical conditions. We generate \Nsyngals synthetic star-forming galaxies composed of up to 3000 \hii regions and explore how variations in parameters  affect their integrated emission-line properties. Our results highlight systematic biases in oxygen-abundance estimates derived using traditional methods, emphasizing the importance of accounting for the composite nature of star-forming galaxies when interpreting integrated nebular emission. Future work will leverage this approach to explore in detail its impact on parameter estimates of star-forming galaxies.
\end{abstract}




\section{Introduction}

Emission lines in the integrated spectra of star-forming galaxies provide valuable clues about the physical conditions in the interstellar medium (ISM) heated by young massive stars, such as gas metallicity, electron density and temperature, ionization conditions and dust content. In practice, interstellar parameters, particularly gas-phase metallicities, of spatially unresolved galaxies have been determined using various emission lines at ultraviolet, optical and infrared wavelengths using different techniques. Common approaches include the `direct method’ based on electron-temperature estimates from auroral lines \citep[e.g.][]{1959Aller_apj130, 1967Peimbert_apj150, 1969Peimbert_Bole5, 2017Perez-Montero_pasp129}, `empirical’ calibrations linking auroral-line metallicities to strong-line flux ratios \citep[e.g.][]{1976Jensen_apj209, 1979Alloin_aap78, 1979Pagel_mnras189, 2008Kewley_apj681, 2013Marino_aap559}, and fully theoretical approaches using photoionization models \citep[e.g.][see \citealt{2023Fernandez_mnra520} for a comparison between the direct method and a model-based method]{2001Charlot_mnra323, 2002Kewley_apjs142, 2016Morisset_aap594, 2018Byler_apj863, 2021Perez-Montero_mnra504}.  Each method has its own limitations and can lead to significant discrepancies between the derived measurements \citep[see the review by][]{2019Kewley_ARAA57}.

Another complication in estimating interstellar parameters from the integrated nebular emission of star-forming galaxies is that these galaxies typically harbour a population of \hii regions with different temperatures, ionization parameters and metallicities. Yet, all the above standard methods are generally applied under the assumption that the ISM conditions in star-forming galaxies can be described by global parameters at the galaxy scale. \citet{1999Kobulnicky_apj514} showed that, while analyses of global emission-line spectra using standard methods provide fairly reliable constraints on chemical properties, they introduce small systematic errors due to varying gas conditions. The pioneering analysis by \citet{1999Kobulnicky_apj514} was based on a limited set of six combinations of two spectra from \hii regions of fixed metallicity sampling three values of electronic temperature and two values of ${\rm O^+/O^{2+}}$. \citet{2014Mast_aap561} explore how spatial resolution affects observed line ratios and derived oxygen abundances, and how integrating the emission of entire galaxies biases spectral analyses. Also relevant is the introduction by \citet{1997Ferguson_apj487}, to model the spectra of Seyfert galaxies, of the concept of `Locally Optimally emitted Cloud', where a distribution of clouds with a wide range of densities and distances to the ionizing source is used to reproduce the narrow-line emission. This concept has been used by \citet{2016Richardson_mnra458} and \citet{2024Lebouteiller_arXi} in the case of star-forming galaxies. 
Recently, \citet{2023Cameron_mnra522} used a radiation-hydrodynamics simulation of an isolated dwarf galaxy to show that temperature fluctuations in the ionized gas throughout the galaxy lead to an underestimation of the true metallicity in estimates of \OH from global emission using the direct method. This type of simulation is, however, very time-consuming and cannot be easily performed for hundreds of thousands of galaxies. Yet, our understanding of the populations of \hii regions in nearby star-forming galaxies has rapidly advanced in recent years, largely thanks to integral-field spectroscopic observations \citep[e.g.][]{2013Marino_aap559, 2018Rousseau-Nepton_mnra477, 2022Santoro_aap658}.

In this paper, we introduce a novel approach to model the nebular emission from star-forming galaxies hosting large populations of \hii regions spanning wide ranges of physical properties. Our approach relies on the efficiency of machine learning (ML) to interpolate an unlimited number of models from a carefully-sampled reference grid encompassing extensive ranges of \hii-region parameters. 

A wide range of ML-based algorithms can be used to perform interpolation in high-dimensional spaces, spanning from basic linear interpolation to more sophisticated techniques. For example, \citet{2018Galliano_mnra476} used linear interpolation to predict dust models, while \citet{2022Ramambason_aap667} applied a nearest-neighbor approach to interpolate within a grid of half a million models. \citet{2019Wu_mnra484} used radial-basis functions to interpolate within a grid of 1367 photodissociation-region models, and \citet{2022Smirnov-Pinchukov_aap666} employed a k-nearest neighbour regressor to interpolate within a grid of one million thermochemical protoplanetary-disc models. Furthermore, \citet{2021Bron_aap645} used random-forest regressors to predict observables in a chemistry-model grid, whereas \citet{2024Maltsev_aap681} leveraged Gaussian-process regression to emulate stellar tracks in
a five-dimensional parameter space. 
Artificial neural networks (ANNs) have also been explored, with \citet{2019Ho_mnra485} employing them to interpolate within a grid of emission-line observations to derive oxygen abundances, \citet{2023Palud_aap678} investigating the chemical composition and emission of diffuse clouds and photodissociation regions, and \citet{2024Li_arXi} developing the \Cue tool to predict stellar and nebular emission using ANNs trained on a set of \cloudy\footnote{\cloudy is a versatile photoionization code designed to simulate the physical and chemical conditions in interstellar matter across of wide range of environments. It computes radiation transfer through a region photoionized by a user-defined spectral energy distribution, assuming thermal and ionization equilibrium. The code provides a variety of outputs, including emission-line intensities, ionic fractions, and electron temperature; see \href{https://nublado.org}{https://nublado.org} for more details.} models \citep{2017Ferland_rmxa53}. 
Autoencoder-based techniques (a subclass of ANNs) have also gained prominence, with \citet{2024Badenas-Agusti_mnra529} applying them to interpolate within a grid of 25,000 stellar-atmosphere models, and  \citet{2021Holdship_aap653} and \citet{2022Grassi_aap668} using autoencoders to emulate thermochemistry models and chemical networks, respectively.  

In this work, we opt for a regression ANN with a simple architecture due to its exceptional speed during the prediction phase, which is crucial for our large-scale production of \hii-region models to be incorporated in synthetic galaxy-emission models.

The method presented in this paper involves developing, training, and testing a regression ANN to predict model properties in record time for any set of input parameters. Artificial neural networks are a class of algorithms with architectures loosely inspired by the biological neural networks found in animal brains. Their adaptability and ability to model non-linear relations make them highly effective for a broad range of tasks, such as classification, regression and clustering, which has contributed to their widespread adoption in scientific fields, in particular in astrophysics. In this study, we use ANNs as regressors: predicting one or more floating-point outputs from a set of floating-point inputs. The ANN model is trained by optimizing its parameters (weights and biases of the individual neuron-like functions) through an iterative process which minimizes the difference between the predicted outputs and the actual outputs in the training set. For more details on the use of machine learning in astrophysics, and a comprehensive description of ANN characteristics, see \citet{2019Baron_arXi}.

We use this method to compute the emission-line properties of \Nsyngals synthetic star-forming galaxies, obtained by combining the contributions from 100 to about 3000 \hii regions per galaxy, incorporating loose trends in physical properties, random dust attenuation, a predefined \Ha luminosity function, and a diffuse gas component ionized by HOt Low-Mass Evolved Stars \citep[HOLMES, see][]{2011Flores-Fajardo_mnra415}. Our method allows exploration of how different parameter distributions affect the integrated emission-line properties of composite galaxies. As an example of application, we study systematic biases in \OH estimates using the direct method, particularly at low and high metallicities. Our results highlight the importance of accounting for the composite nature of star-forming galaxies when interpreting integrated nebular emission, a topic we will explore further in the next paper in this series.

We present the reference grid of photoionization models of \hii regions in Section~\ref{sec:grid} and the ANN and its performance in Section~\ref{sec:ML}. In Section~\ref{sec:synthetic_galaxies}, we describe our approach to computing the nebular emission from synthetic star-forming galaxies. The properties of these galaxies are presented in Section~\ref{sec:results}, together with our study of the systematic biases implicit in \OH estimates using the direct method. Section~\ref{sec:conclusion} summarizes our conclusions.

\section{An original grid of nebular-emission models of \hii\ regions}
\label{sec:grid}

The high dimensionality of the parameter space we wish to explore prevents us from adopting a regular Cartesian grid, the realization of which would require unmanageable resources. We therefore adopt a different approach, which consists in: (i) running an irregular grid of photoionization models, where the different adjustable parameters can take random values drawn from selected distributions; (ii) using machine-learning (ML) techniques to train, based on this grid, a model able to predict emission-line ratios of interest for given input values of the adjustable parameters; and (iii) generating in record time, with this ML model, grids of $10^8$ or more models for any (regular or not) distribution of adjustable parameters. The use of an irregularly-sampled model grid for the training set has a number of advantages. It avoids oversampling of single parameter values with no influence on certain observables, minimizes potential discretization effects in interpolation, and allows the grid size to be increased more easily than with a Cartesian grid.

In this section, we describe our approach to producing a comprehensive grid of models. We start by defining the set of adjustable parameters in Section~\ref{sec:sub:irregulargrids}, with particular attention to chemical abundances and ionizing spectra. Then, in Section~\ref{sub:testhii}, we define the distributions of these parameters.

\subsection{Irregular model grid in high-dimensional parameter space}
\label{sec:sub:irregulargrids} 

We compute photoionization models of \hii regions using the multi-purpose photoionization code \cloudy in its version 17.03. They are defined, run and stored under the reference `HII\_24' in 3MdB\_17,\footnote{3MdB is a large database of photoionization models, with public access through the mySQL protocol since 2014 \citep[\url{http://3mdb.astro.unam.mx}, see][]{2015Morisset_rmxa51}.} a database using the \pycloudy facility \citep{2014Morisset_}. 

The adjustable parameters of the \cloudy models considered in this work are: 
\begin{itemize}
    \item the hydrogen density \nh;
    \item the gas filling factor $\epsilon$;
    \item the ionization parameter $U$: unless otherwise specified, $U$ refers here to the mean ionization parameter over the volume $V$ of the H$^+$ region, $U=\bar U= \frac{1}{V}\int_V U(r) 4\pi r^2\mathrm{d}r$, with $U(r)= Q_0/(4\pi r^2\nh c)$ at radius $r$. In this expression, $Q_0$ is the rate of ionizing photons, and $c$ is the speed of light. For a nebula with constant \nh confined between R$_1$ and R$_2$ (as in the models considered here), $\bar U = 3U(R_2)(1-R_1/R_2)/[1-(R_1/R_2)^3]$;
    \item the form factor \fr, defined as the ratio $f_\mathrm{r}=R_\mathrm{in}/R_\mathrm{S}$ of the model inner radius to the Str\"omgren radius corresponding to $R_\mathrm{in}=0$. This parameter characterizes the shape of the nebula, transitioning from fully spherical for $f_\mathrm{r}\ll1$ to plane-parallel for $f_\mathrm{r}\gg1$ \citep[see section~4.1 of ][where this parameter is called $f_\mathrm{S}$]{2015Stasinska_aap576}. The relation between $\bar U$, $Q_0$, \nh, $\epsilon$ and \fr for a pure-hydrogen nebula is: 
 $ Q_0 = 4 \pi c^3 \bar U^3 / \{ 3 \alpha_B^2 n_H \epsilon^2 [(f_\mathrm{r}^3 + 1)^{1/3} - f_\mathrm{r}]^{3} \}$, where $\alpha_B$ is the case-B recombination coefficient;
    \item the chemical composition, i.e. the abundances relative to hydrogen of the 28 heavy elements handled by \cloudy (see Section~\ref{sub:abunds});
    \item the parameter \ksid, as defined by \citet{2016Gutkin_mnra462}, which represents the mass fraction of heavy elements depleted on to dust grains (see Section~\ref{sub:abunds});
    \item the parameter \HbFrac: each individual \cloudy model generates ten entries in the database, after being sliced at different optical depths corresponding to matter-bounded cases. The depth of the cut is parametrized by the quantity \HbFrac, defined as the ratio of the \Hb luminosity of the truncated model to that of the radiation-bounded model. Thus $\HbFrac<1$ corresponds to matter-bounded models and $\HbFrac=1$ to radiation-bounded models \citep[see section~4.1 of][]{2015Stasinska_aap576}. The parameter \HbFrac can be approximately related to \fesc, the fraction of ionizing photons escaping from the nebula, by the expression $\fesc = 1 - \HbFrac$ (we also consider below, in Section~\ref{sec:sub:gal_param_distribution}, the `picket-fence' model with gas covering factor less than unity, where photons can escape through holes in the nebula);
    \item the age $t_\mathrm{c}$ and metallicity $Z$ of the ionizing star cluster, which determine the shape of the ionizing spectrum (see Section~\ref{sub:SEDs} for more details). In this work, we adopt the same metallicity for the stars as for the gas they ionize. 
\end{itemize}

\subsubsection{Abundances, depletion and dust}

\label{sub:abunds}

We start by drawing the total (i.e. in both the gas and dust phases) oxygen abundance of a model, \OHtot. The abundances of all other chemical elements are drawn relative to \OHtot, following the prescriptions of \citet[][their table~2]{2017Nicholls_mnra466}. We update the values of \NHtot and \CHtot by randomly drawing from the chosen distributions of \NOtot and \COtot (Section~\ref{sub:testhii}). In this approach, \NOtot, \COtot and \OHtot are therefore not correlated with one another, nor are they with the ionization parameter, which we draw fully independently. Any relation between these quantities in galaxies should therefore arise from comparisons of models with observations. In fact, the results of \citet{2020Schaefer_apjl890} show the importance of not imposing a fixed relation between, for example, \NOtot and \OHtot in emission-line studies of star-forming galaxies. We also perturb all other metal abundances around their initial values. Once a complete set of total abundances is obtained, we compute the associated metallicity $Z$, which we take to be the same for the ionizing star cluster.

We account for the depletion of all refractory elements on to dust grains according to the values in table~2 of \citet{2013Dopita_apjs208}, scaled in such a way as to match the randomly-drawn dust-to-metal mass ratio, \ksid. The resulting set of gas-phase abundances is used as input for \cloudy. In the remainder of this paper, all abundances refer to the gas phase, except where indicated by a superscript `tot'. Finally, the dust-to-gas ratio (DTG) entered in \cloudy is determined as $\mathrm{DTG/DTG_\odot} = \ksid Z/(\ksidsol \Zsun)$, with $\ksidsol = 0.36$ and $\Zsun = 0.01425$. For simplicity, we use the `ism' dust type of \cloudy with argument $\mathrm{DTG/DTG_\odot}$, meaning that there no strict coherence between element-by-element depletion and dust composition.

\subsubsection{Stellar ionizing spectra}
\label{sub:SEDs}

The models presented in this study rely on a set of ionizing spectral energy distributions computed using the BPASS~v2.2\footnote{\url{https://bpass.auckland.ac.nz/}} model of binary-star populations \citep[][]{2018Stanway_mnra479, 2022Byrne_mnra512}. We adopt a fixed \citet{2003Chabrier_apjl586} initial mass function (IMF) with lower and upper cutoffs 0.1 and 300~M$_\odot$, respectively.

We note that although some recent studies have suggested that X-ray-binary accretion discs may significantly contribute to the ionizing radiation in metal-poor star-forming galaxies \citep{2019Schaerer_aap622, 2022Umeda_apj930, 2023Katz_mnras518}, models calibrated on the observed average luminosity function of X-ray binaries in nearby galaxies do not support this hypothesis \citep[][see also \citealt{2020Senchyna_mnra494}]{2024Lecroq_mnras527}. We ignore this contribution in our analysis.

\subsection{Production of H II-region models}
\label{sub:testhii}

\begin{table}
\caption{Adjustable parameters used to generate the 3MdB/`HII\_24' grid of \cloudy \hii region models, along with their distributions defined by the minimum and the maximum values of a `plateau' of constant probability, and the width $\sigma_w$ of the Gaussian function used to describe the drop of probability on both sides of this plateau. A random shift $\Delta\log\XHtot$ is applied to the total abundance of any metal X (other than C, N and O) relative to the value defined in the adopted prescription (Section~\ref{sub:abunds}). The distributions refer to total abundances (i.e. including elements in both the gas and dust phases), the dust-to-metal mass ratio \ksid allowing transformation from these to the gas-phase values considered in all figures (see text for details).}
\label{tab:3mdb_params}
\centering
\begin{tabular}{lccc}
\hline
Parameter & min & max & $\sigma_w$ \\
\hline
\logU & $-4$ & $-1$ & 0.5 \\
12+$\log\OHtot$ & 6.5 & 9.5 & 0.5 \\
$\log\NOtot$ & $-2$ & 0 & 0.5 \\
$\log\COtot$ & $-1$ & 0.5 &  0.5 \\
$\Delta\log\XHtot$ &  $-0.1$ & 0.1 & 0.0 \\
$t_\mathrm{c}$/Myr &  1 &  7 & 0.0 \\
$\log f_\mathrm{r}$ & $-1.5$ &  1.5 &  0.0 \\
$\log(n_\mathrm{H}/\mathrm{cm}^{-3})$ & 1 & 4 & 0.5 \\
$\log \epsilon$ & $-2$ & 0 & 0 \\
\ksid & 0.15 & 0.55 & 0.0 \\
\HbFrac &\multicolumn{3}{c}{from 0.1 to 1.0 in steps of 0.1} \\
\hline
\end{tabular}
\end{table}

The grid of \hii-region models presented here is based on the set of adjustable parameters described in the previous sections. We draw values of the adjustable parameters from distributions defined in most cases by three variables: the minimum and the maximum values of a `plateau' of constant probability, and the width $\sigma_w$ of the Gaussian function used to describe the drop of probability on both sides of this plateau. The values controlling the parameter distributions are listed in Table~\ref{tab:3mdb_params}, where we indicate in each case the range and the width $\sigma_w$: some distributions have smooth edges, parametrized by the width $\sigma_w=0.5$ of the Gaussian distribution defining the wings (Section~\ref{sec:sub:irregulargrids}). The value $\sigma_w = 0$ corresponds to a distribution strictly limited to the plateau range. We adopt quasi-log-uniform distributions for most parameters to effectively cover broad value ranges spanning multiple orders of magnitude, whereas the parameter \HbFrac is sampled uniformly between 0.1 and 1.0 in steps of 0.1.

The \COtot and \NOtot abundance ratios are allowed to vary in great proportions, while the other metals are only slightly perturbed (within $\pm0.1\,$dex) around the predefined values obtained from the \citet{2017Nicholls_mnra466} prescription (Section~\ref{sub:abunds}). We allow this to explore how changes in the abundance of any single element can affect observables. 

The 3MdB database currently holds, under the reference `HII\_24', about 3.5~million models
drawn from the distributions listed in Table~\ref{tab:3mdb_params}. These were obtained from 400,000 runs of \cloudy, as each run produces up to ten entries in the database once the optical depth cuts are applied (Section~\ref{sec:sub:irregulargrids}). The models include both families of BPS and BPB ionizing spectra, corresponding to BPASS single- and binary-star populations (Section~\ref{sub:SEDs}). We primarily focus in this paper on results obtained with \NHIIBPB BPB models. 

\begin{figure*}
\includegraphics[width=18cm]{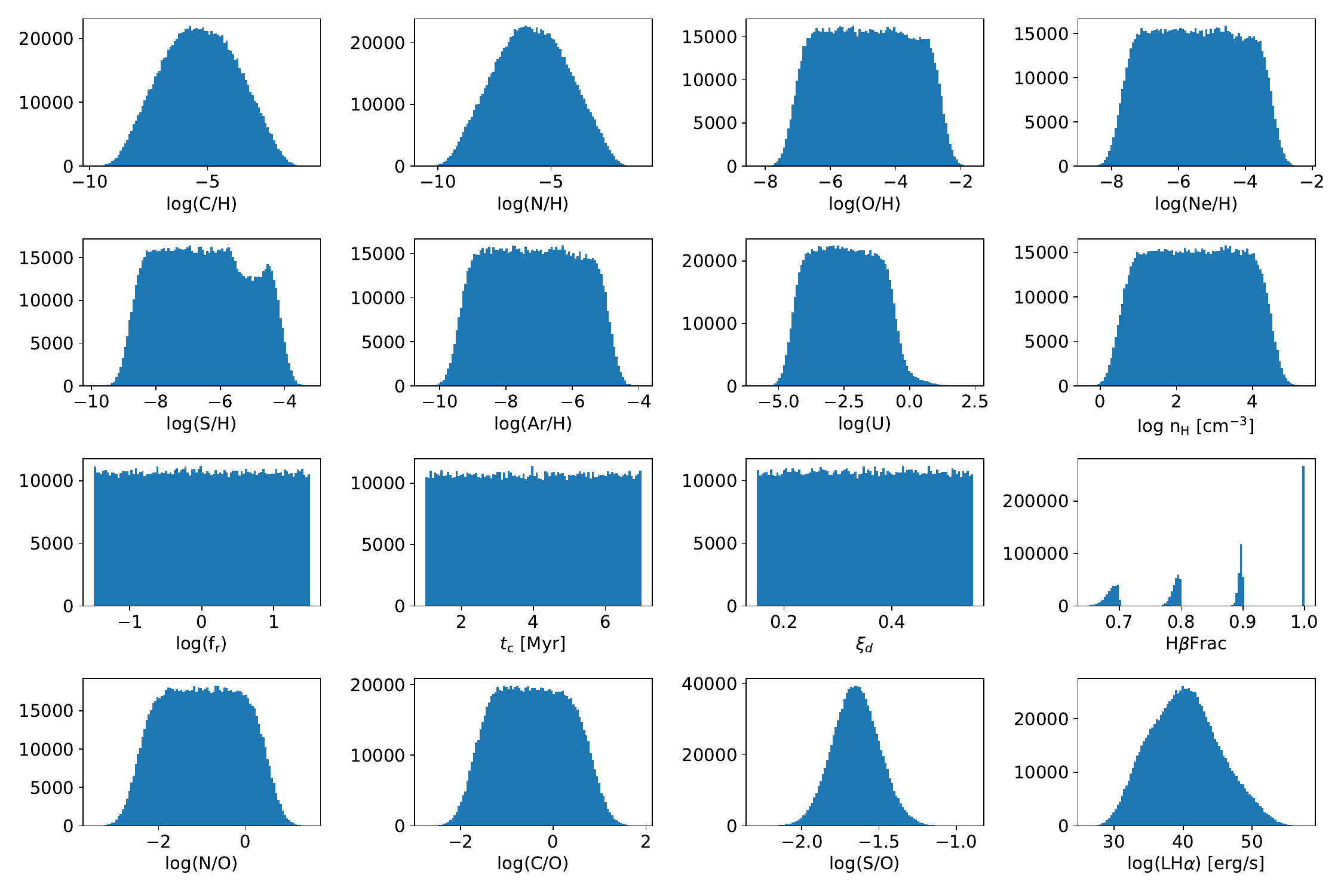}
\caption{Distributions of selected input parameters of one million \cloudy models with $t_\mathrm{c}<7\,$Myr and $\HbFrac > 0.65$ produced using the distributions in Table~\ref{tab:3mdb_params}. Also shown in the bottom right-hand corner is the distribution of output \Ha luminosities.
}
\label{fig:parameters_distribution_3mdb}
\end{figure*}

In Fig.~\ref{fig:parameters_distribution_3mdb}, we plot the distributions of various quantities obtained for one million BPB models with $t_\mathrm{c}<7\,$Myr and $\HbFrac > 0.65$ (corresponding to $\fesc\la0.35$; Section~\ref{sec:sub:irregulargrids}). As expected, distributions with $\sigma_w = 0$ (e.g. for the age $t_\mathrm{c}$) exhibit sharp edges, and those with $\sigma_w = 0.5$ (e.g. for $\log \nh$) smooth edges. The distributions of element abundances are in general not entirely regular, as they are defined indirectly. Indeed, the distributions in Table~\ref{tab:3mdb_params} pertain to total (gas+dust phase) abundances, while those shown in Fig.~\ref{fig:parameters_distribution_3mdb} refer to gas-phase abundances, affected by the dust-to-metal mass ratio \ksid drawn on a case-by-case basis.

The small drop between $-5.5$ and $-4.5$ in the $\log(\mathrm{S/H})$ distribution results from the change in  slope of the relation linking \SHtot to \OHtot according to \citet[][see their fig.~8]{2017Nicholls_mnra466}. There is no similar drop in the distributions of Ar/H and Ne/H, as the relations defining the abundances of these elements relative to oxygen do not exhibit any change in  slope \citep[table~2 of][]{2017Nicholls_mnra466}. We note the small tails of \HbFrac-distribution spikes below the discrete values drawn from Table~\ref{tab:3mdb_params}. This is because we do not interpolate between `zones' (radial steps) of computation defined by \cloudy and only integrate the \Hb luminosity out to the zone ending closest to the selected \HbFrac. Similarly, the \logU distribution shows a small tail toward large values, since the distribution in Table~\ref{tab:3mdb_params} pertains to the radiation-bounded model, while the volume-averaged ionization parameter tends to increase toward the interior of the nebula and is then higher in the matter-bounded models (\HbFrac < 1.0).

The simplest way to populate a multi-dimensional grid is to assign parameters in each dimension to regularly-spaced values, either in linear or logarithmic scales, as used by e.g. \citet{2016Morisset_aap594}, \citet{2016Vale-Asari_mnra460}, \citet{2018Galliano_mnra476}, \citet{2022Ramambason_aap667} and \citet{2022Smirnov-Pinchukov_aap666}. For extremely large grids, a more practical approach involves using randomly-distributed independent parameters, which offers an efficient way to populate datasets to train ML algorithms \citep[e.g.][]{2021Bron_aap645}. 
In future iterations of our method to generate a grid of \hii regions, we plan to adopt more efficient distributions of input parameters. This will help mitigate random clustering of data points in certain regions of the parameter space, which could introduce biases into the training of the ANN. Specifically, we aim to explore methods such as Latin Hypercube Sampling and quasi-random sampling techniques like Sobol and Halton sequences, as used by, e.g., \citet{2022Mootoovaloo_Astr38}, \citet{2019Bird_jcap2019} and \citet{2022Reza_}, as well as the method described by \citet{korobov1959approximate} and applied by, e.g., \citet{2022DeRose_jcap2022} and \citet{2024Ramirez_jcap2024}. With such approaches, the plateau-like parameter distributions with smoothly decreasing edges used in the present work will no longer be preserved.

\section{Regression based on machine learning to interpolate nebular-emission models}
\label{sec:ML}

In this section, we use machine learning to interpolate unlimited amounts of models from the grid presented in Section~\ref{sec:grid}. We present the regression model, evaluate its performance and explore a few applications by examining the influence of the main adjustable parameters on selected observables. Our approach consists in creating, training and testing a regression ANN, here managed through the AI4Neb library\footnote{\label{foot:AI4Neb}AI4Neb allows the trained network to be saved, along with the scaling function (used to normalize the input data) and the training and test sets), as described by Morisset et al. (in preparation). See \href{https://github.com/Morisset/AI4neb}{https://github.com/Morisset/AI4neb}} to predict the properties of models with any given input parameters. The neural network counts three hidden layers with 256, 512 and 256 activation cells, respectively. All activation functions are of the Gaussian Error Linear Unit (GELU) type \citep{2016Hendrycks_arXi}. A detailed description of the ANN is provided in Appendix~\ref{sec:app:ANN}.

\subsection{Regression model}
\label{sec:sub:regressor}

We use \NHIItrain models for the training set (a random selection of 80~per cent of the models in Fig.~\ref{fig:parameters_distribution_3mdb}) to train an ANN receiving 13 parameters as input,

\begin{itemize}
    \item $\log U$
    \item $\log f_\mathrm{r}$
    \item $\log t_\mathrm{c}$
    \item $\log($O/H)
    \item $\log($C/O)
    \item $\log($N/O)
    \item $\log($Ne/O)
    \item $\log($S/O)
    \item $\log($Ar/O)
    \item $\log n_\mathrm{H}$
    \item H$\beta$Frac
    \item $\xi_{\rm{d}}$
    \item $\epsilon$\,,
\end{itemize}
and predict, as output, 14 line-luminosity ratios, the electronic temperature of the ionized gas and the \Ha luminosity (all in logarithmic scale), 
\begin{itemize}
    \item H$\alpha$/\Hb
    \item {[O~{\sc iii}]$\lambda$5007/\Hb}
    \item {[N~{\sc ii}]$\lambda$6584/\Hb} 
    \item {[O~{\sc ii}]$\lambda$3726,3729/\Hb (hereafter [O~{\sc ii}]$\lambda$3727/\Hb)}
    \item {[O~{\sc iii}]$\lambda$4363/$\lambda$5007}
    \item {[N~{\sc ii}]$\lambda$5755/$\lambda$6584}
    \item {[S~{\sc ii}]$\lambda$6716,6731/\Hb (hereafter [S~{\sc ii}]$\lambda$6725/\Hb)}
    \item {[S~{\sc iii}]$\lambda$9530/\Hb}
    \item {[Ar~{\sc iii}]$\lambda$7135/\Hb}
    \item {[Ne~{\sc iii}]$\lambda$3868/\Hb}
    \item {[S~{\sc ii}]$\lambda$6716/$\lambda$6731}
    \item He~{\sc i}$\lambda$5876/\Hb
    \item {[S~{\sc iii}]$\lambda$6312/$\lambda$9530}
    \item {[O~{\sc ii}]$\lambda$7320,7330/\Hb (hereafter [O~{\sc ii}]$\lambda$7325/\Hb)}
    \item {$T_\mathrm{e}$}
    \item $\LHa$\,.
\end{itemize}

The list of inputs is in practice redundant, as, for example, S/O is fully determined once O/H and \ksid are known. Such redundancy is commonly employed to enhance the performance of regression models by providing extra information strongly related with the desired outputs -- in this case, the  luminosities of S emission lines \citep[see, e.g.,][where redundant line ratios and sums are used]{2019Ho_mnra485}.

We have included the electronic temperature as an output variable, because it is treated as such (and not as an input parameter) in our \cloudy models and will enable exploration of the influence of temperature fluctuations on the integrated line emission from a collection of \hii regions (Section~\ref{sec:synthetic_galaxies}). Also, the \Ha luminosity is a key output quantity to identify models which may result from unlikely combinations of randomly drawn input parameters.

\subsection{Performance of the regression model}
\label{sec:sub:quality_regressor}

\begin{figure*}
\includegraphics[width=18cm]{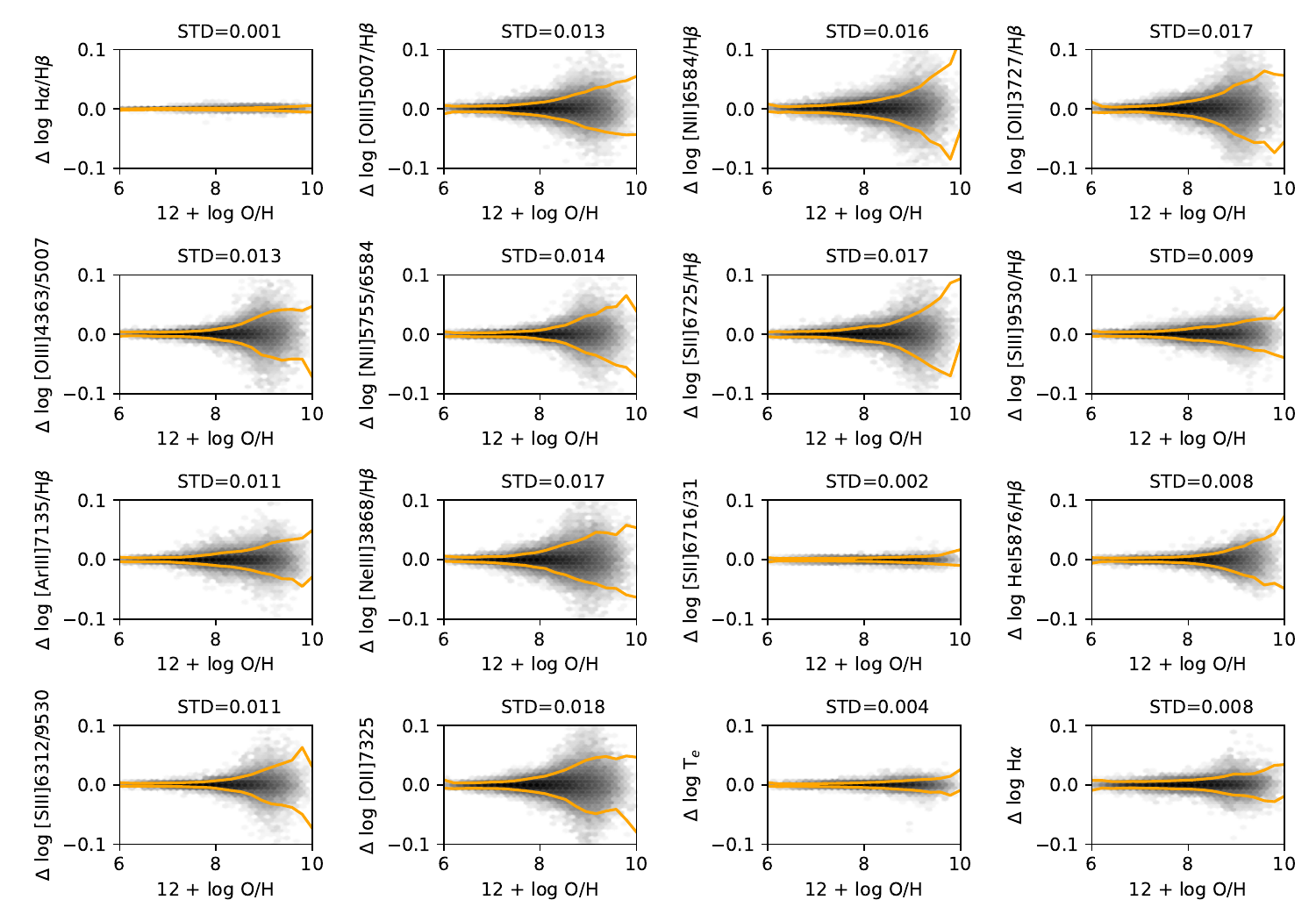}
\caption{Density maps of the difference between true and ANN-predicted values (both in logarithm) of 14 emission-line ratios, the electronic temperature and the absolute H$\alpha$ luminosity, plotted against \logOH, for the 20 per cent of models from Fig.~\ref{fig:parameters_distribution_3mdb} not included in the ANN training set (darker colour corresponding to higher model density). In each panel, the orange lines show the fifth-95th percentile range of the distribution as a function of \logOH, while the standard deviation of the differences between true and predicted logarithmic values is indicated at the top.}
\label{fig:fit_quality2}
\end{figure*}

\begin{figure*}
\includegraphics[width=18cm]{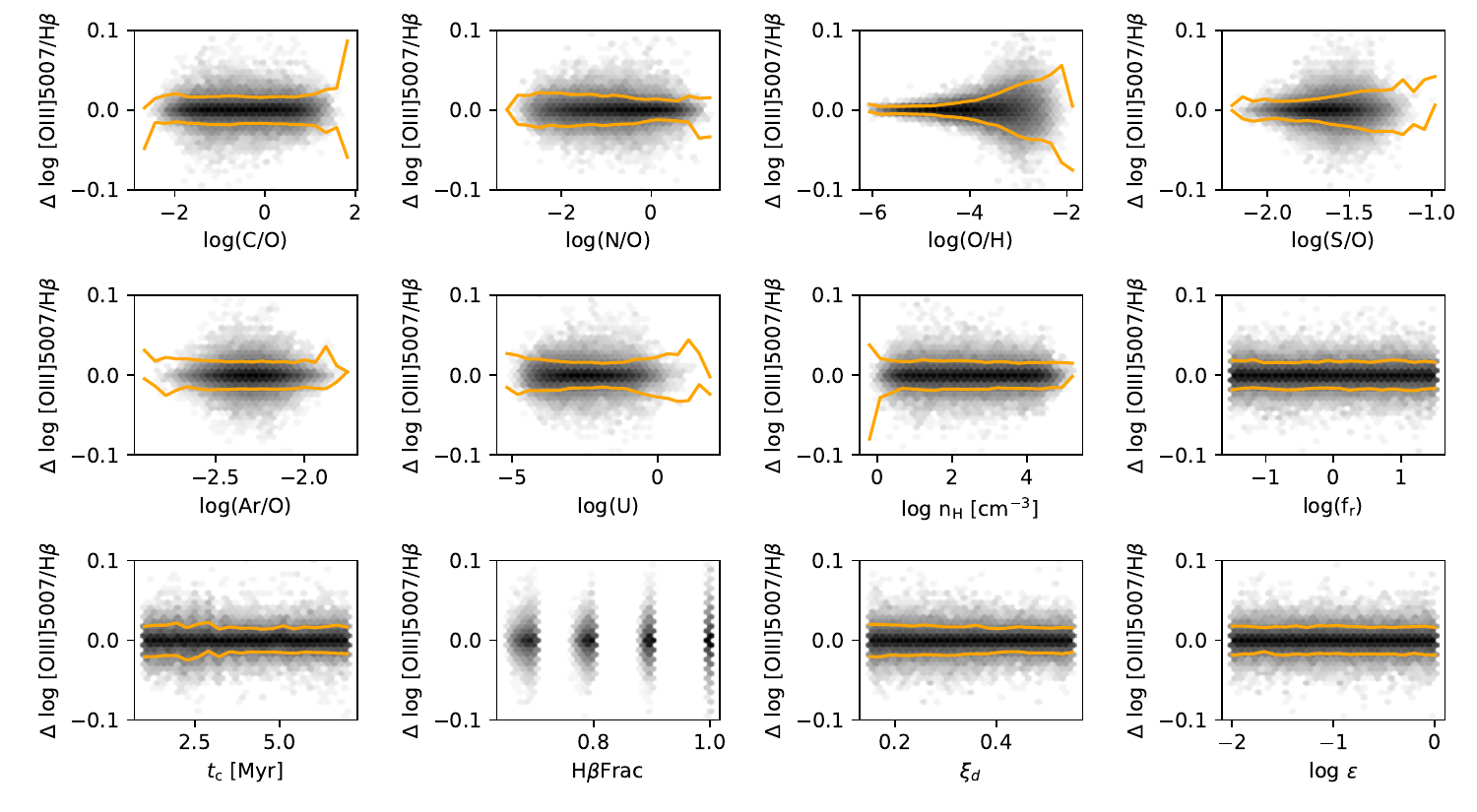}
\caption{Analogue of Fig.~\ref{fig:fit_quality2} for the logarithmic difference between true and predicted \forba{O}{iii}{5007}/\Hb ratio as a function of each of the 13 input parameters of the ANN of Section~\ref{sec:sub:quality_regressor}.}
\label{fig:fit_quality3}
\end{figure*}

\begin{figure}
\includegraphics[width=8cm]{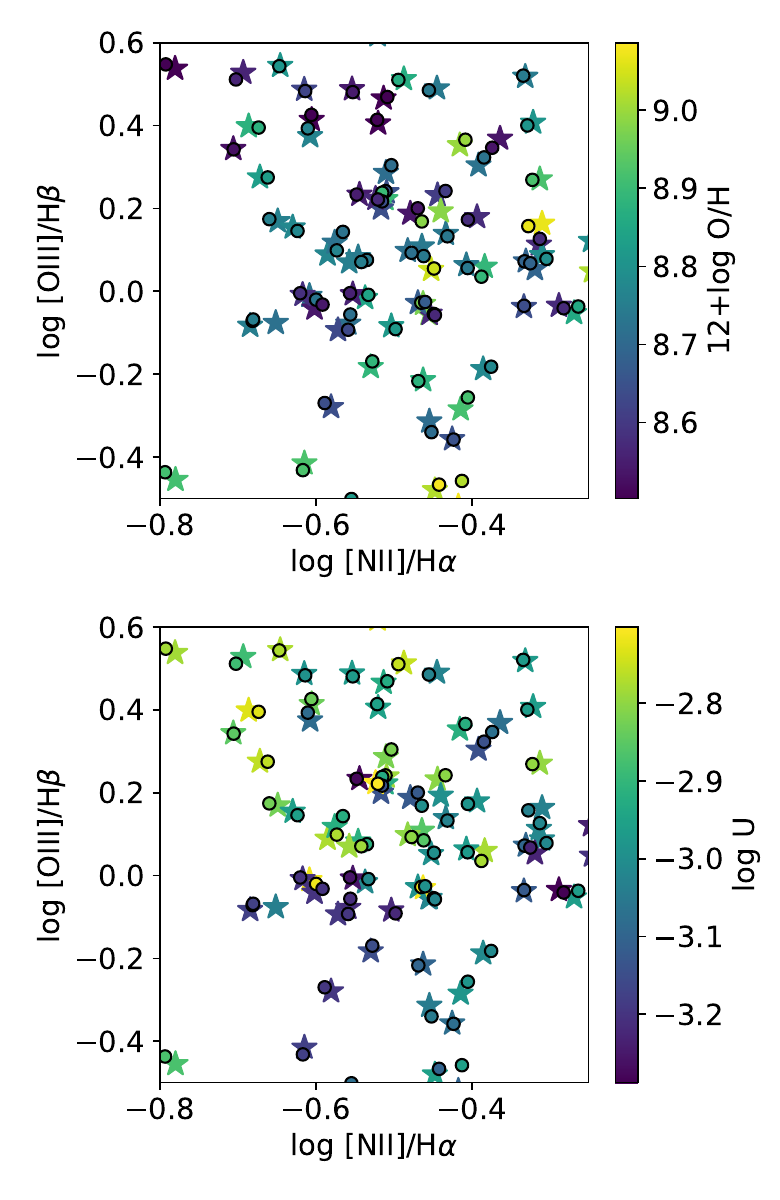}
\caption{True (stars) and predicted (circles) line ratios in the \forba{O}{iii}{5007}/\Hb-versus-\forba{N}{ii}{6584}/\Ha (BPT) diagram, for a random subset of 40 models from the sample shown in Fig.~\ref{fig:fit_quality2}. The models are colour-coded according to oxygen abundance (left) and ionization parameter (right).}
\label{fig:fit_quality}
\end{figure}

Fig.~\ref{fig:fit_quality2} shows the distributions of the differences between the true values of all 14 line ratios and the electronic temperature (all in logarithmic scale) and the values of these quantities retrieved from the regression model. To evaluate the quality of the ANN, we focus on the subset of 200,000 test-set points projecting into a specific triangular region of the \forba{O}{iii}{5007}/\Hb-versus-\forba{N}{ii}{6584}/\Ha `BPT' diagram \citep{1981Baldwin_pasp93a}. This region corresponds to the location of the models used in subsequent steps and is defined by the following three conditions: $\forba{O}{iii}{5007}/\Hb < 10$; $\forba{N}{ii}{6584}/\Ha < 1$; and $\log(\forba{O}{iii}{5007}/\Hb) > -1.1\log(\forba{N}{ii}{6584}/\Ha) - 2.6$. 
Applying this filter selects 61,110 data points from the 200,000 available in the test set.
Distributions are mapped as a function of oxygen abundance, with orange lines indicating the fifth-95th percentile range. We also indicate, at the top of each panel, the global standard deviation of the difference between true and retrieved values. To complement Fig.~\ref{fig:fit_quality2}, we show in Fig.~\ref{fig:fit_quality3} the difference between true and predicted log(\forba{O}{iii}{5007}/\Hb) ratio as a function of each of the 13 input parameters of the regression model. Finally, in Fig.~\ref{fig:fit_quality}, we compare true (stars) and predicted (circles) line ratios for a random subset of 40 models in the BPT diagram, colour-coded according to oxygen abundance (left) and ionization parameter (right).

Figs~\ref{fig:fit_quality2}--\ref{fig:fit_quality} together demonstrate the high quality of the regression model. The standard deviation in Fig.~\ref{fig:fit_quality2} is less than 0.02~dex for all quantities, in stark contrast to the scatter in corresponding observed values, which can exceed 1\,dex in datasets such as the SDSS (see Fig.~\ref{fig:Olaws} discussed in the next section). Moreover, the errors in the predictions are relatively symmetric, being equally distributed between positive or negative differences. This reinforces our confidence in the regression model, as small symmetric errors can compensate for each other when exploring trends with a large number of predicted models. In Fig.~\ref{fig:fit_quality}, true and predicted models are also very close to each other, with no systematic effect. Figs~\ref{fig:fit_quality2} shows a clear trend of increasing errors toward large O/H: at metallicities above $\logOH\approx 9$, the regression model starts to lose accuracy. At such high metallicities, the low electron temperature amplifies the impact of small parameter variations, which can be more difficult to track with the ANN model. Such metallicities are barely reached in the applications presented in the remainder of this paper.   

We conclude that the regression ANN presented in this section provides an efficient means to predict the observable properties of individual \hii regions. It allows us to generate a large number of photoionization models and explore the influence of input parameters on emission-line predictions much faster than would be required with \cloudy.

\begin{table}
\centering
\caption{Performance of the regression ANN in predicting emission-line properties of \hii regions across various CPU and GPU configurations. The first column lists the number of models run by the ANN, while the second and third columns indicate the number of CPUs and GPUs used. The fourth column shows the peak RAM usage for CPUs, no data being given when GPU is used. The last column provides execution time.}
\begin{threeparttable}

\begin{tabular}{ccccc}
\toprule
N models &CPU\tnote{a} & GPU\tnote{b} & max memory used [GB] &  time [s]\\
\midrule
    & 1 & 0 & 2.8 &  2.08 \\
    & 12 & 0 & 2.8 & 0.27 \\
2.0$\times 10^5$ & 12\tnote{c} & 0 & 2.8 & 0.65 \\
    & 12\tnote{d} & 0 & 2.8 & 1.8 \\
    & 0 & 1 & -- & 0.036 \\
\hline
    & 1 & 0 & 16 & 12 \\
    & 12  & 0 & 16 & 1.75 \\
    & 12\tnote{c}  & 0 & 16 & 9.6 \\
1.6$\times 10^6$ & 12\tnote{d} & 0 & 16 & 36\tnote{e} \\
    & 60 & 0 & 16 & 1.15 \\
    & 0 & 1 & -- & 0.26 \\
\hline
   & 1 & 0 & 123 &  131 \\
   & 12 & 0 & 123 & 13.2 \\
1.3$\times 10^7$ & 30 & 0 & 123 & 8.8 \\
   & 60 & 0 & 123 & 8.8 \\
\bottomrule
\end{tabular}
\begin{tablenotes}
\item [a] AMD EPYC 7313 processor, unless otherwise specified.
\item [b] NVIDIA RTX A5000 graphics card.
\item [c] M2 Pro processor on a MacMini.
\item [d] Intel Core i7 2.6\,GHz processor on a MacBook Pro.
\item [e] In this configuration, the RAM was filled, triggering a swap process which increased the execution time.
\end{tablenotes}
\label{tab:CPUGPU-perf}
\end{threeparttable}
\end{table}

Our ANN can reliably predict (Figs~\ref{fig:fit_quality2}--\ref{fig:fit_quality} and Table~\ref{tab:ANNs}) the electronic temperature, \Ha luminosity, and strengths of 14 line ratios for $10^6$ combinations of the 13 input parameters, corresponding to as many \hii regions, in less than 0.2 seconds using a single GPU (i.e. nine orders of magnitude faster than running \cloudy on a single CPU). Details on the performance of the model predictions, including memory and time usage for different grid sizes, are provided in Table~\ref{tab:CPUGPU-perf}, comparing CPU and GPU configurations. 

We note that the ANN can also be used as part of an algorithm to search for the parameters that best reproduce observations of a given \hii region (such as Markov-Chain Monte Carlo and genetic algorithms). This will be the subject of a forthcoming paper.

\subsection{Exploring the influence of input parameters on emission-line predictions for \hii regions}
\label{sec:sub:hii_preds}

\begin{figure*}
\includegraphics[width=18cm, trim={0 0 5cm 0},clip]{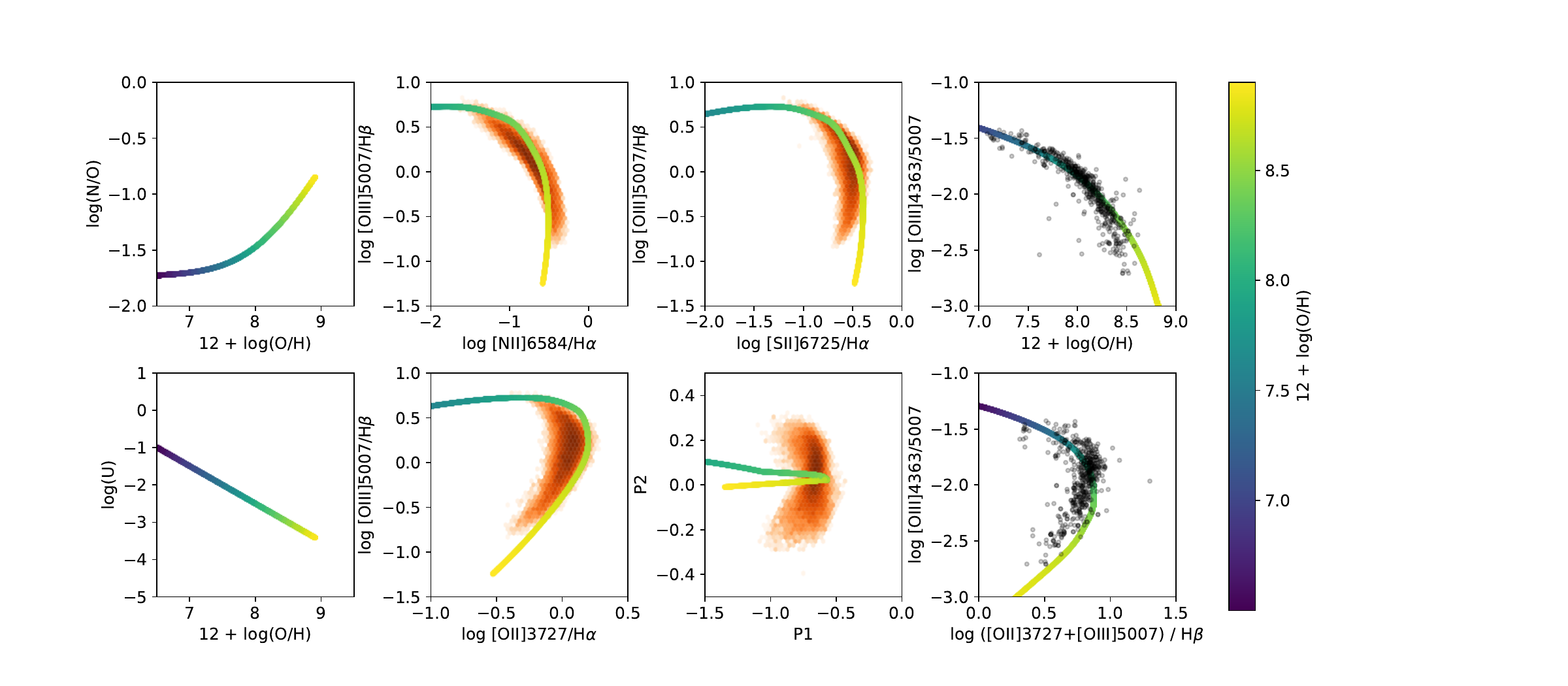}
\caption{Properties of 10,000 regression-ANN models with both N/O and $U$ tied to O/H (according to the relations shown in the leftmost panels) in six diagrams: the \forba{O}{iii}{5007}/\Hb-versus-\forba{N}{ii}{6584}/\Ha (BPT) diagram; the \forba{O}{iii}{5007}/\Hb-versus-\forba{S}{ii}{6725}/\Ha and \forba{O}{iii}{5007}/\Hb-versus-\forba{O}{ii}{3727}/\Ha diagrams (VO87); the $P_2$-versus-$P_1$ diagram defined by \citet{2022Ji_aap659}; and two diagrams showing the temperature-sensitive \forba{O}{iii}{4363}/5007 ratio against \logOH (top right) and the (\forba{O}{ii}{3727} + \forba{O}{iii}{5007})/\Hb ratio (bottom right). The models have fixed ionizing-cluster age, $t_\mathrm{c}=3\,$Myr, hydrogen density, $\nh=30\,$cm$^{-3}$, form factor, $f_\mathrm{r}=1.0$, $\HbFrac=0.95$ and dust-to-metal mass ratio $\ksid=0.35$. The models are colour-coded according to oxygen abundance (see scale on the right). The orange symbols (in the four central panels) show observations of SDSS star-forming galaxies, and the black symbols (in the rightmost panels) observations of \hii regions selected from the SDSS-MaNGA survey (see text for more details).}
\label{fig:Olaws}
\end{figure*}

With the ANN developed to enable fast modelling of \hii regions, we can efficiently explore how basic relations between input parameters, such as $\log U$ versus O/H, impact on the predicted observables in classical diagnostic diagrams. This analysis has didactic value, as the results can be directly compared with survey data, to provide insight into the validity and quality of particular assumptions. In this section, we explore some of these relations.

The four central panels of Fig.~\ref{fig:Olaws} show the locations of 10,000 models with both N/O and $U$ tied to O/H (according to the relations drawn in the leftmost panels) in four line-ratio diagrams: the \forba{O}{iii}{5007}/\Hb-versus-\forba{N}{ii}{6584}/\Ha (BPT) diagram; the \forba{O}{iii}{5007}/\Hb-versus-\forba{S}{ii}{6725}/\Ha and \forba{O}{iii}{5007}/\Hb-versus-\forba{O}{ii}{3727}/\Ha diagrams \citep[][hereafter VO87]{1987Veilleux_apjs63}; and the $P_2$-versus-$P_1$ diagram defined by \citet{2022Ji_aap659} by combining \Ha, [\ion{N}{ii}], [\ion{S}{ii}]  and [\ion{O}{iii}] lines.\footnote{\label{foot:plines}$P_1 = 0.63 \forba{N}{ii}{6584}/\Ha + 0.51 \forba{S}{ii}{6725}/\Ha + 0.59 \forba{O}{iii}{5007}/\Hb $ and $P_2 = -0.63 \forba{N}{ii}{6584}/\Ha + 0.78 \forba{S}{ii}{6725}/\Ha$.} The models have fixed ionizing-cluster age, $t_\mathrm{c}=3\,$Myr, hydrogen density, $\nh=30\,$cm$^{-3}$, form factor, $f_\mathrm{r}=1.0$, $\HbFrac=0.95$ and dust-to-metal mass ratio $\ksid=0.35$. Also shown in the four central panels of Fig.~\ref{fig:Olaws} are the observed properties of \NSDSS star-forming (i.e. \texttt{BPTCLASS}=1) SDSS galaxies from the DR8 MPA-JHU catalogue,\footnote{\url{https://www.sdss4.org/dr17/spectro/galaxy_mpajhu/}} selected to have signal-to-noise ratio $\mathrm{S/N}>10$ in all strong lines of interest and $\mathrm{S/N}>3$ in \alloa{He}{i}{5876} and \forba{Ne}{iii}{3869} (orange background). In the rightmost panels, we compare the models to observations (in black) of 603 \hii regions compiled by \citet{2013Marino_aap559} in two diagrams showing the temperature-sensitive \forba{O}{iii}{4363}/5007 ratio against \logOH (top) and the (\forba{O}{ii}{3727} + \forba{O}{iii}{5007})/\Hb ratio (bottom).

In all panels of Fig.~\ref{fig:Olaws}, the models follow a narrow sequence set by the unique relation between O/H, N/O and $U$ chosen for the purpose of this illustration. The sequence overlaps with at least some observations in all line-ratio diagrams, but does not reflect the observed scatter. In the $P_1$--$P_2$ diagram, unlike in other diagrams, the models do not reproduce at all the observational sequence, particularly along the $P_2$ axis. The sequence is captured when N/O or \logU are allowed to deviate from the fixed relation, as discussed below. 

\begin{figure*}
\includegraphics[width=18cm, trim={0 0 5cm 0},clip]{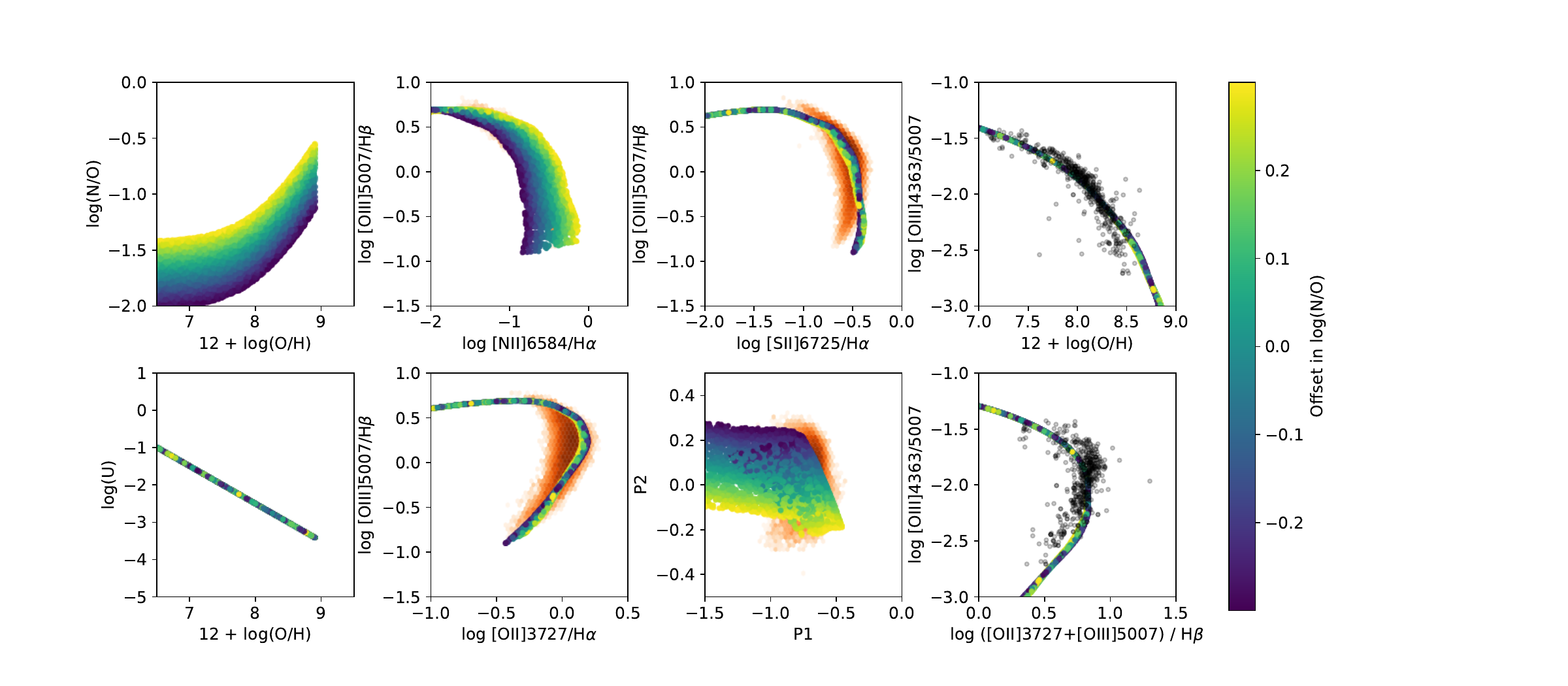}
\includegraphics[width=18cm, trim={0 0 5cm 0},clip]{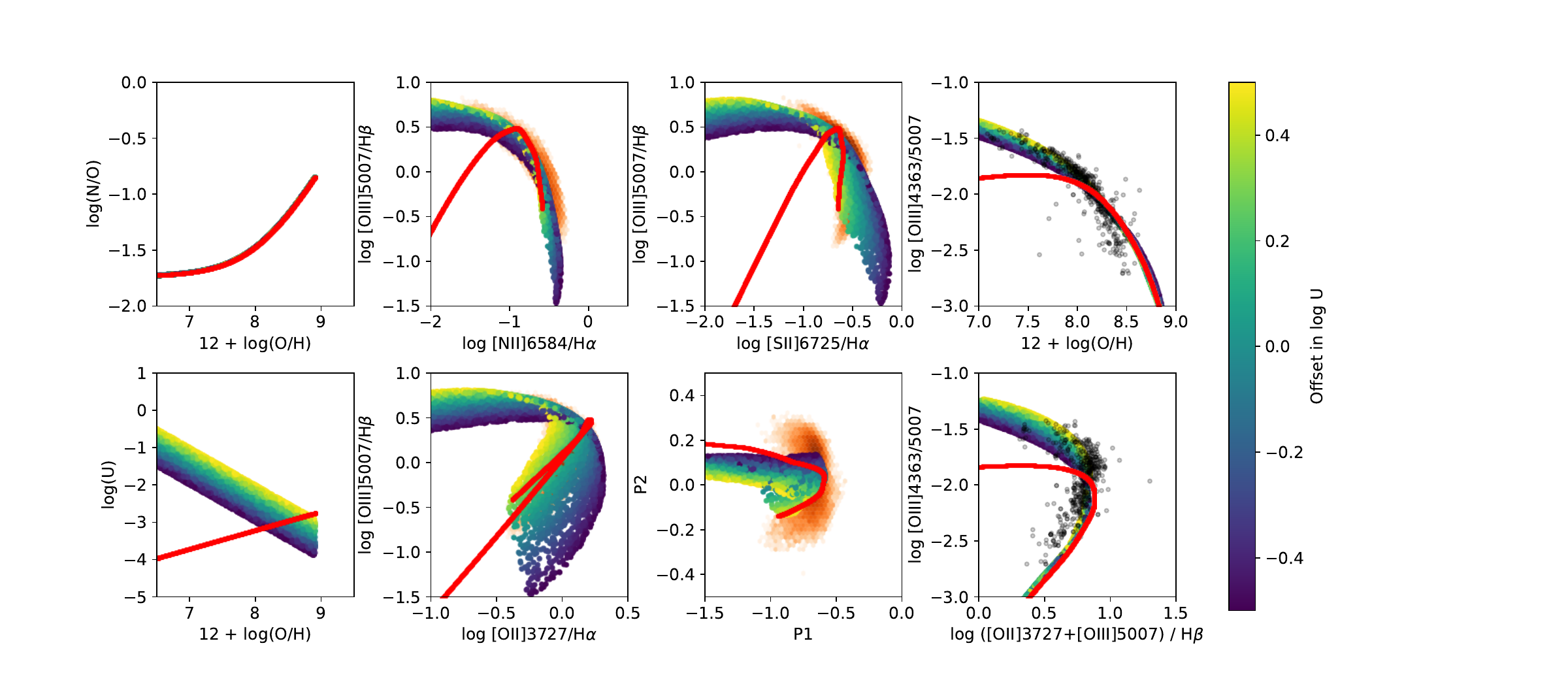}
\caption{The top eight panels show the analogue of Fig.~\ref{fig:Olaws} for the same 10,000 models, but drawing N/O uniformly withing $\pm0.3$~dex of the relation in the top-left panel of Fig.~\ref{fig:Olaws} (as indicated by the colour scale on the right). The bottom eight panels show the analogue of Fig.~\ref{fig:Olaws} for the same 10,000 models, but drawing $U$ uniformly withing $\pm0.5$~dex of the relation in the bottom-left panel of Fig.~\ref{fig:Olaws} (as indicated by the colour scale on the right). Also shown in red are the results for 10,000 models with $\logU = -7.2 + 0.5[\logOH]$ \citep[following][]{2022Ji_aap659}. The data are the same as in Fig.~\ref{fig:Olaws}.}
\label{fig:Olaws2}
\end{figure*}

We explore the effect of allowing for variations in log(N/O) and \logU at fixed \logOH in Fig.~\ref{fig:Olaws2}. The top eight panels show the results for 10,000 models as in Fig.~\ref{fig:Olaws}, but drawing N/O uniformly withing $\pm0.3$~dex of the relation in the top-left panel of Fig.~\ref{fig:Olaws}. The bottom eight panels of Fig.~\ref{fig:Olaws2} show the results for 10,000 models as in Fig.~\ref{fig:Olaws}, but drawing $U$ uniformly withing $\pm0.5$~dex of the relation in the bottom-left panel of Fig.~\ref{fig:Olaws}. For completeness, in these panels we also draw 1000 models for which the ionizing parameter is set to increase with oxygen abundance, as found by \citet[][their fig.~10]{2022Ji_aap659} for extragalactic \hii regions selected from the SDSS-MaNGA survey, which we parametrize roughly as $\logU = -7.2 + 0.5[\logOH]$ (in red).

Fig.~\ref{fig:Olaws2} illustrates how modest changes in the adopted dependence of N/O and $U$ on O/H can significantly affect the predicted line ratios. The influence of changes in N/O at fixed O/H is strong on ratios involving [\ion{N}{ii}] lines, but negligible on other line ratios. In contrast,  changes in the dependence of $U$ on O/H significantly affect all line ratios, including the $T_\mathrm{e}$-sensitive \rforba{O}{iii}{4363}{5007} ratio. Such variations allow the models to cover a larger portion of the observational sequences in the various diagrams. We note that the model where \logU increases with \logOH leads to predictions far from the observations at low metallicity.This positive correlation identified by \citet{2022Ji_aap659} may result from their use of a fixed relation between N/O and O/H, following the prescription of \citet[][]{2013Dopita_apjs208}.

In Appendix~\ref{sec:app:HIIregion} we explore the effect on line ratios of changing the other adjustable parameters of the models, namely: ionizing cluster age, hydrogen density, \HbFrac and  dust-to-metal mass ratio. 

\section{Building the nebular emission from composite star-forming galaxies}
\label{sec:synthetic_galaxies}

The ANN presented in the previous section allows us to calculate in record time the emission-line properties of large numbers of \hii\ regions with arbitrary physical parameters. In this section, we exploit this tool to build synthetic star-forming galaxies by efficiently generating and assembling large populations of \hii\ regions with different parameters, including individual gas-covering factors. These regions are individually attenuated by dust, and a component of diffuse ionized gas (DIG) is added to the composite galaxy. Then, we apply a global correction for dust attenuation to the emission-line luminosities of these synthetic galaxies using the \Ha/\Hb\ ratio, following the approach typically employed for observed galaxies. We describe the detailed steps of this process in the following subsections.

\subsection{Distribution of \hii-region parameters}
\label{sec:sub:gal_param_distribution}

We compute the nebular emission from composite star-forming galaxies by summing the emission-line luminosities from a collection of \hii regions. The emission lines considered here are all optically thin and are affected only by dust attenuation, which is accounted for by assigning a randomly determined A$_V$ value to each \hii region (Section~\ref{sec:sub:diffuseISM}).

For each galaxy model, we start by drawing a `baseline' galactic oxygen abundance, \OHgal, around which we draw the oxygen abundances of individual \hii regions, \OH. We write 
\begin{equation}
    \logOHgal = 8  + \varrho(\delta_\mathrm{O/H})\,,
\label{eq:OH}
\end{equation}
where $\varrho(\delta_\mathrm{O/H})$ is a uniformly distributed random function taking values between $-\delta_\mathrm{O/H}$ and $+\delta_\mathrm{O/H}$, and
\begin{equation}
    \log(\OH) = \log\OHgal + \varrho(\sigma_\mathrm{O/H})\,.
\label{eq:OH2}
\end{equation}
Hence, the quantity $\delta_\mathrm{O/H}$ defines the departure of the baseline O abundance from the canonical $12+\log\OHgal=8$ value, while the quantity $\sigma_\mathrm{O/H}$ defines the random variation of \OH around this baseline abundance from one \hii region to another, in a given galaxy. 

We define other nebular parameters in a similar way, by drawing baseline galactic values through standard relations of the parameter of interest to O/H, and \hii-region values by sampling around these baseline galactic values. Specifically, we assume that the ionization parameter scales roughly as the inverse of metallicity \citep{2017Carton_mnra468}, adopting a canonical value $\log U_\mathrm{gal}=-2.5$ at the canonical O abundance $\logOHgal = 8$ (equation~\ref{eq:OH}). We thus write
\begin{equation}
\begin{split}    
    \log U = -2.5 - [\log(\OH) + 4] +
    \varrho(\delta_\mathrm{U}) + \varrho(\sigma_\mathrm{U})\,.
\end{split}    
\label{eq:logU}
\end{equation}
For the dependence of \CO, \NO, \NeO, \SO and \ArO on \OH, we start from the baseline relations derived by \citet[][their table~2]{2017Nicholls_mnra466} for total abundances, which we convert into relations for gas-phase abundances via the dust-to-metal mass ratio \ksid tied to the depletion factors of \citet[][see Section~\ref{sub:abunds} above]{2013Dopita_apjs208}. We achieve this by computing, for a grid of \OHtot and \ksid sampling the full ranges of these quantities in Table~\ref{tab:3mdb_params}, the corresponding gas-phase abundances \OH, \CO, \NO, \NeO, \SO, and \ArO. We use this data set to train a small artificial neural network (called abund-ANN) to interpolate \CO, \NO, \NeO, \SO and \ArO based on \OH and \ksid.\footnote{The abund-ANN is developed using the Keras library (\href{https://github.com/keras-team/keras}{https://github.com/keras-team/keras}) running on the Tensorflow  back end (\href{http://tensorflow.org/}{http://tensorflow.org/}). It consists of two hidden layers, each with 50 neurons using GELU activations \citep[][]{2016Hendrycks_arXi}, trained with the Adam optimizer \citep{2014Kingma_arXi} for 200 epochs on a training set of 32,000 models. The input is of dimension two and the output of dimension five. The standard deviations of the differences between the predictions and the test values are of the order of 0.002. The abund-ANN is managed through the AI4Neb library.} With this framework, for each galaxy for which a baseline galactic \OH was selected from equation~\eqref{eq:OH2}, we select a baseline dust-to-metal mass ratio uniformly in the range $0.2\leq\ksid\leq0.5$ and use the abund-ANN to determine the corresponding abundance ratios $(\XO)_\mathrm{fit}$, where X stands for N, Ne, S or Ar. By analogy with equations~\eqref{eq:OH}--\eqref{eq:OH2}, we define baseline abundance ratios for this galaxy as
\begin{equation}
  \log\XOgal = \log(\XO)_\mathrm{fit} + \varrho(\delta_\mathrm{X/O})\,,
\label{eq:XOgal}
\end{equation}
and the abundance ratios of individual \hii regions as
\begin{equation}
  \log(\XO) = \log(\XO)_\mathrm{gal} + \varrho(\sigma_\mathrm{X/O})\,.
\label{eq:XO}
\end{equation}
We note that the quantities $\varrho(\delta_\mathrm{U})$ and $\varrho(\delta_\mathrm{N/O})$ in equations~\eqref{eq:logU}--\eqref{eq:XOgal} above provide a means of investigating small changes in the adopted dependence of N/O and $U$ on O/H, whose importance was noted in Section~\ref{sec:sub:hii_preds} (Fig.~\ref{fig:Olaws2}). The quantity $\varrho(\delta_\mathrm{N/O})$ can be assimilated to the nitrogen excess factor considered by \citet[][]{2020Schaefer_apjl890} and denoted by $\xi$ in their nomenclature.

\begin{table*}
\caption{Adjustable parameters used to generate composite star-forming galaxies (see Section~\ref{sec:sub:gal_param_distribution} for details).}
\label{tab:PySynGal_params}
\begin{threeparttable}
\centering
\begin{tabular}{lccl}
\hline
\multicolumn{1}{l}{Parameter} & \multicolumn{1}{c}{Galaxy spread} & \multicolumn{1}{c}{\hii-region spread} & \multicolumn{1}{c}{Description}\\
\hline
$\log\mathrm{(O/H)}$ & $\delta_\mathrm{O/H}$ = 1.2 & $\sigma_\mathrm{O/H}$=0.15 & Oxygen abundance (see equations~\ref{eq:OH}--\ref{eq:OH2} for canonical value)\\
$\log U$ & $\delta_\mathrm{U}$ = 0.75 & $\sigma_\mathrm{U}$=0.5 & Ionization parameter (canonical value from equation~\ref{eq:logU})\\
\ksid & [0.20, 0.50] & $\sigma_{\ksid}$=0.05 & Dust-to-metal mass ratio\\
$\log\mathrm{(C/O)}$ & $\delta_\mathrm{C/O}$ = 0.3 & $\sigma_\mathrm{C/O}$=0.2 & C/O abundance ratio (canonical value tied to O/H and \ksid; Section~\ref{sec:sub:gal_param_distribution})\\
$\log\mathrm{(N/O)}$ & $\delta_\mathrm{N/O}$ = 0.3 & $\sigma_\mathrm{N/O}$=0.2 & N/O abundance ratio (canonical value tied to O/H and \ksid; Section~\ref{sec:sub:gal_param_distribution})\\
$\log\mathrm{(S/O)}$ & $\delta_\mathrm{S/O}$=0.3 & $\sigma_\mathrm{S/O}$=0.2 & S/O abundance ratio (canonical value tied to O/H and \ksid; Section~\ref{sec:sub:gal_param_distribution}) \\
$\log\mathrm{(Ne/O)}$ & $\delta_\mathrm{Ne/O}$=0.3 & $\sigma_\mathrm{Ne/O}$=0.2 & N/O abundance ratio (canonical value tied to O/H and \ksid; Section~\ref{sec:sub:gal_param_distribution}) \\
$\log\mathrm{(Ar/O)}$ & $\delta_\mathrm{Ar/O}$=0.3 & $\sigma_\mathrm{Ar/O}$=0.2 & Ar/O abundance ratio (canonical value tied to O/H and \ksid; Section~\ref{sec:sub:gal_param_distribution}) \\
\HbFrac & [0.75, 0.95] & $\sigma_\mathrm{HbFrac}$=0.05 & \Hb-luminosity ratio of truncated versus radiation-bounded models\\
\DIGFrac  & [0.0, 0.4] & N/A & \Hb-luminosity ratio of DIG versus \hii regions\\
$t_\mathrm{c}^\mathrm{min}$/Myr & [1, 7]& N/A & Minimum ionizing star-cluster age\\
$t_\mathrm{c}^\mathrm{max}$/Myr & [$t_\mathrm{c}^\mathrm{min}$, 7] & N/A & Maximum ionizing star-cluster age\\
$t_\mathrm{c}$/Myr & N/A & [$t_\mathrm{c}^\mathrm{min}$, $t_\mathrm{c}^\mathrm{max}$]& \hii-region age\\
$\log(n_\mathrm{H}/\mathrm{cm}^{-3}$) & N/A & $[0.5, 3]$& Hydrogen density\\
log $\epsilon$& N/A & $[-2, 0]$ & Volume filling factor of the gas\\
A$_V$/mag  & N/A & [0, 2] & $V$-band dust attenuation\\
$\log\fr$ & N/A & $[-1.5, 1.5]$ & Geometric form factor\\
\fc & N/A & [0.1, 1.0] & Gas covering factor \\
\aLF & N/A & N/A & Slope of the \Ha luminosity function of \hii regions, fixed at $-1.9$\\
\Nregmin & N/A & N/A & Minimum number of \hii regions, fixed at 100\\

\hline
\end{tabular}
\end{threeparttable}
\end{table*}
In Table~\ref{tab:PySynGal_params}, we list the spread in baseline galactic values (second column), and that in individual \hii-region values around a given baseline value (third column), for the oxygen abundance, ionization parameter and the different metal-abundance ratios explored here (top seven lines). To probe the influence of star formation history, we randomly define for each galaxy the minimum and maximum \hii-region ages, $t_\mathrm{c}^\mathrm{min}$ and $t_\mathrm{c}^\mathrm{max}$, within the range of cluster ages used to train the ANN (Table~\ref{tab:3mdb_params}). We then randomly draw the ages of all the galaxy's \hii regions between $t_\mathrm{c}^\mathrm{min}$ and $t_\mathrm{c}^\mathrm{max}$. This allows us to include galaxies containing either only very young bursts, only older bursts, or a broad distribution of burst ages, as might be the case, for example, with constant star formation. For the form factor \fr, hydrogen density $n_\mathrm{H}$, volume-filling factor of the gas $\epsilon$ and dust-to-metal mass ratio \ksid, we sample over the full ranges used to train the ANN, while we restrict \HbFrac to values in the range 0.7--1.0, corresponding to (relatively) radiation-bounded \hii regions.

To simulate star-forming galaxies as realistically as possible, we must also specify a luminosity distribution of \hii regions such that the \Ha luminosity function. Firstly, we account for the fact that, in addition to being potentially matter-bounded, \hii regions can also lose ionizing photons through holes resulting from incomplete covering of the ionizing source by the surrounding gas. Following \citet{2011Heckman_apj730}, we parametrize such a `picket-fence’ model in terms of the fraction $\fc=\Omega/4\pi$ of solid angle over which the gas covers the ionizing star cluster, which we randomly draw between 0.1 and 1.0 for each \hii region. The effect is that all line luminosities are reduced by a factor \fc, without affecting the line ratios.

Then, for the \Ha luminosity function of \hii regions, we adopt a power-law distribution $\mathrm{LF} (\LHa) \propto \LHa^{\aLF}$, where $\mathrm{LF}(\LHa)d\LHa$ is the number of regions with luminosities between \LHa and $\LHa+d\LHa$. We fix the slope at $\aLF=-1.9$, intermediate between the values $-2.12\pm0.03$ and $-1.73\pm0.15$ found from spatially-resolved spectroscopy of $\sim4300$ \hii regions in the nearby, face-on spiral galaxy NGC\,628 \citep{2018Rousseau-Nepton_mnra477}\footnote{The numerical values of the quantity $\alpha$ quoted in \citet{2018Rousseau-Nepton_mnra477} refer to the slope of the logarithmic \Ha luminosity function, $\alpha=\aLF+1$ (L. Rousseau-Nepton, private communication).} and $\sim23,000$ \hii regions in a sample of 19 nearby, low-inclination spiral galaxies \citep{2022Santoro_aap658}. 

In practice, we build the emission-line properties of a composite star-forming galaxy by: (i) drawing the physical parameters of a large sample of \hii regions using the procedure described above (Table~\ref {tab:PySynGal_params}); (ii) using the ANN to compute the associated emission-line properties of these \hii regions (Section~\ref {sec:sub:regressor}); (iii) reducing the emission-line luminosities by the gas covering factor selected for each \hii region;  and (iv) including individual \hii regions in the galaxy emission-line budget or rejecting them on the basis of their \Ha luminosity, as determined by the target luminosity function. We accomplish this last step by associating to each \hii region a chance 
\begin{equation}
    \mathcal{C}= 
\begin{cases}
    \frac{(\LHa/ \LHa_\mathrm{min})^{\aLF}}{\upsilon(0,1)}& \text{if } 
    \LHa_\mathrm{min} <\LHa < \LHa_\mathrm{max},\\
    0              & \text{otherwise,}
\end{cases}
\label{eq:PropaIncluded}
\end{equation}
where $\upsilon(0,1)$ is a random variable drawn for each \hii region from a uniform distribution over the interval from 0 to 1. The exact values of $\LHa_\mathrm{min}$ and $\LHa_\mathrm{max}$ have little influence on our results. We adopt $\LHa_\mathrm{min}= 1\times10^{36}$\,erg\,s$^{-1}$ and $\LHa_\mathrm{max}= 3\times10^{40}$\,erg\,s$^{-1}$ \citep{2018Rousseau-Nepton_mnra477, 2022Santoro_aap658}. An \hii region is included in the galaxy emission-line budget only if the chance given by equation~\eqref{eq:PropaIncluded} exceeds unity, i.e. $\mathcal{C}>1$.

We compute in this way the emission-line properties of \Nsyngals composite star-forming galaxies containing at least $\Nregmin=100$ individual \hii regions. Each galaxy is built by drawing an initial number of 50,000 \hii regions, the majority of which are rejected based on the above criterion on $\mathcal{C}$, so that model galaxies in practice contain up to about 3000 \hii regions. The low efficiency of the process is compensated for by the speed of the ANN. A higher efficiency could be achieved by choosing a different set of input parameters for the regression model (Section~\ref{sec:sub:regressor}), for example, by replacing the ionization parameter and form factor with the ionizing-photon rate and inner radius of the \hii region. The disadvantage is that this would not allow exploration of the influence of the relation between ionization parameter and oxygen abundance (equation~\ref {eq:logU}) on the diagnostic diagrams described in Section~\ref{sec:sub:hii_preds}.

\subsection{Components linked to molecular clouds and the diffuse ISM}
\label{sec:sub:diffuseISM}

We combine the emission from \hii regions in composite star-forming galaxies with components linked to molecular clouds and the diffuse ISM.

To reflect the diversity of molecular-cloud environments in which stars form, we attenuate the emission from each individual \hii region using the Galactic extinction curve of \citet{1999Fitzpatrick_pasp111} with R$_V=3.1$, scaled by a $V$-band dust attenuation, A$_V$, drawn randomly between 0 and 2 (Table~ \ref{tab:PySynGal_params}). This is justified by the fact that, while A$_V$ for a given region can be influenced by its size and metallicity, it also strongly depends on the presence of dust clouds along the line of sight, which is further affected by the orientation of the host galaxy and the position of the region within it. The specific wavelength dependence of dust attenuation has a negligible influence on the results presented in this study.

We also include a component of diffuse ionized gas (DIG), which is an important ingredient of nebular emission in star-forming galaxies \citep[e.g.][]{2017Sanders_apj850, 2019Stasinska_arXi}. For this, we call on the \cloudy calculations of \citet{2023Martinez-Paredes_mnra525}, which describe the emission of interstellar gas photoionized by populations of HOLMES \citep[see][]{2011Flores-Fajardo_mnra415}. These stars are expected to be the main sources of ionizing photons producing the DIG \citep{2022Rautio_aap659, 2023Postnikova_arXi}. We do not consider here the potential (presumably minor) contribution from leakage of ionizing photons from \hii regions \citep[e.g.][]{2024McClymont_mnras}.

We use a regression artificial neural network, called DIG-ANN, similar to that described in Section~\ref{sec:ML} above, to interpolate the DIG emission-line luminosities for each composite galaxy from the \citet{2023Martinez-Paredes_mnra525} model grid. We add a fraction \DIGFrac of the emission-line luminosities obtained in this way, drawn uniformly in the range $0.0\leq \DIGFrac\leq0.4$, to the attenuated line emission from \hii regions, to obtain the total line emission of the composite galaxy. The DIG-ANN and its training set are described in details in Appendix~\ref{sec:app:DIG-ANN}.

Then, we correct the total line emission of the composite galaxy based on the departure of the integrated \Ha/\Hb ratio from the reference dust-free value of 2.86 \citep[appropriate for a gas of density 100\,cm$^{-3}$ and temperature 10,000\,K;][]{2006Osterbrock_}, using the Galactic extinction curve of \citet{1999Fitzpatrick_pasp111}, as would be done for an observed galaxy. We refer to the integrated, dust-corrected \Ha luminosity of the galaxy computed in this way as \LHagal.

\begin{figure*}
\includegraphics[width=18cm]{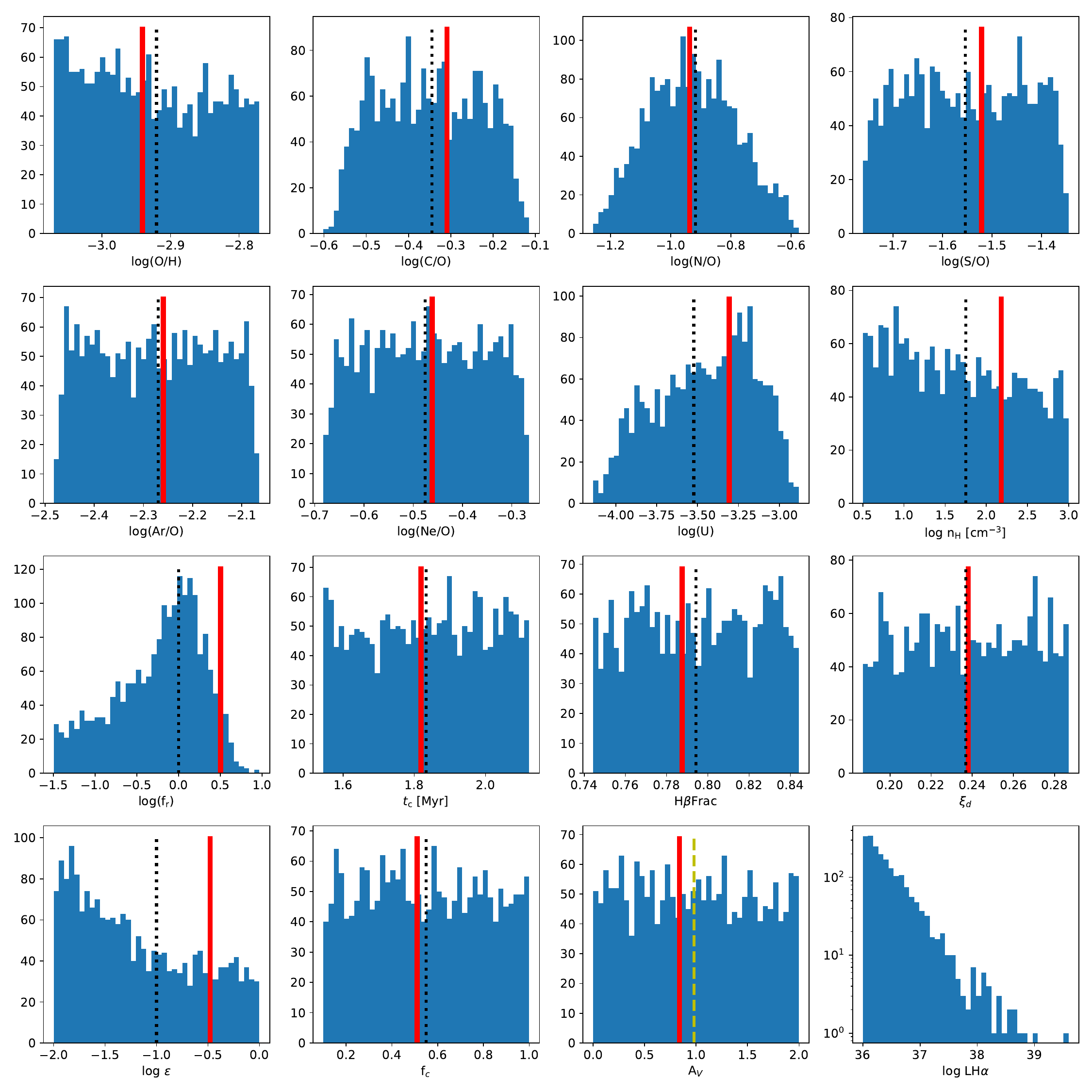}
\caption{Distributions of individual parameters of the 2002 \hii regions constituting one of the composite star-forming galaxies built as described in Section~\ref{sec:sub:gal_param_distribution} (the meaning of the different parameters is described in Table~\ref{tab:PySynGal_params}). The corresponding `baseline' parameters are shown as vertical black dashed lines. The vertical red lines show the average, \Ha-weighted \hii-region parameters  computed as described in Section~\ref{sub:integproperties}. The dashed yellow line in the A$_V$ panel shows the value of this parameter recovered from the global \Ha/\Hb ratio, as described in Section~\ref{sec:sub:diffuseISM}.}

\label{fig:parameters_distribution_oneGal}
\end{figure*}

As an example, we show in Fig.~\ref{fig:parameters_distribution_oneGal} the distributions of individual parameters of the 1616 \hii-regions constituting one such composite star-forming galaxy, for which the baseline ionization parameter and abundances (Section~\ref{sec:sub:gal_param_distribution}) are indicated by vertical dashed lines.

\section{Results}
\label{sec:results}

In this section, we explore the integrated properties of the \Nsyngals composite star-forming galaxies built as described in the previous section. We investigate the location of these galaxies in standard line-ratio diagnostic diagrams. Then, to illustrate the usefulness of this sample to interpret the integrated emission-line properties of composite star-forming galaxies, we examine the extent to which determinations of the oxygen abundance based on the integrated line emission of a galaxy differ from the average abundance of its constituent \hii regions.

\subsection{Integrated properties of composite star-forming galaxies}
\label{sub:integproperties}

\begin{figure*}
\includegraphics[width=18cm]{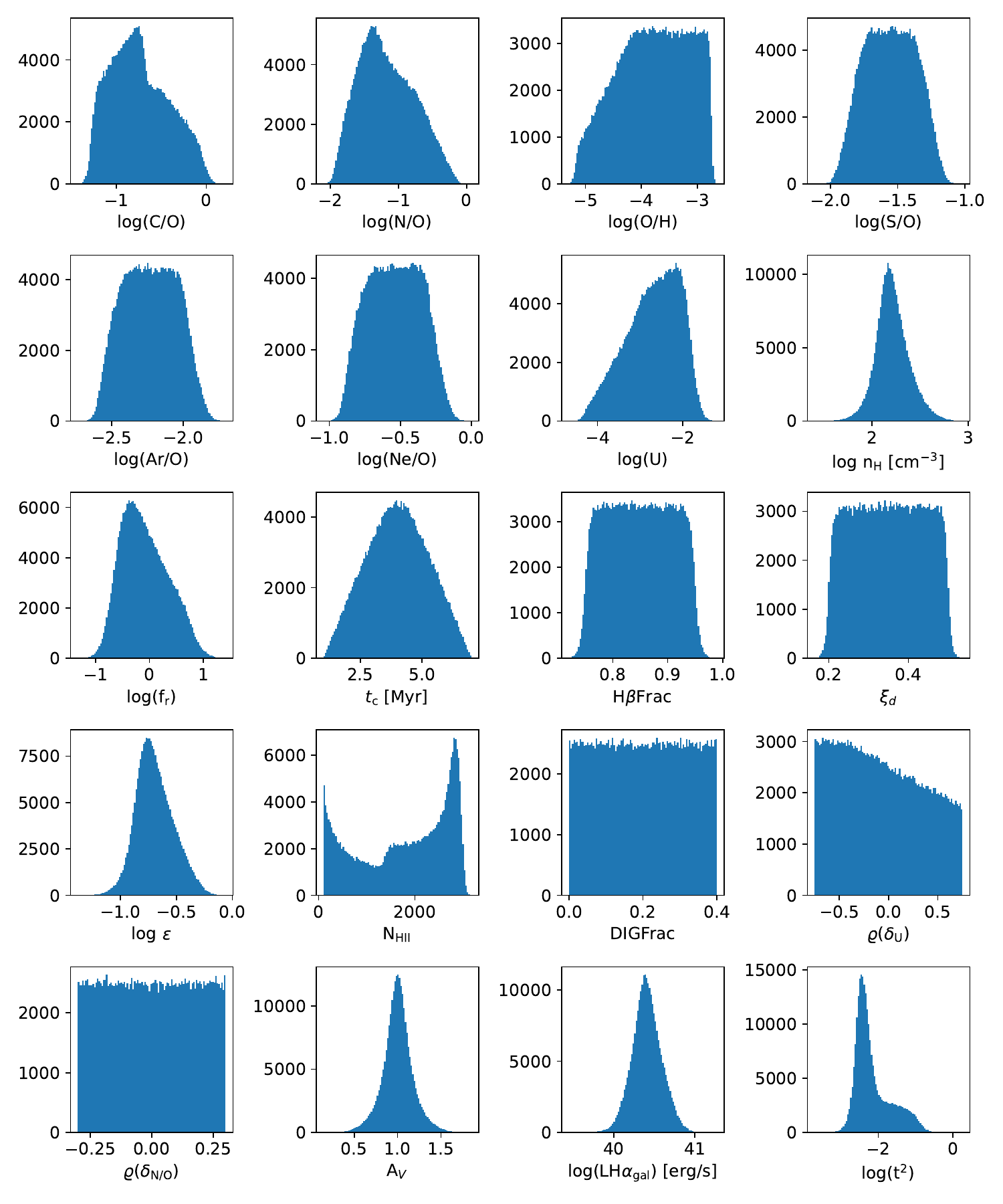}
\caption{Distributions of various properties of the \Nsyngals composite star-forming galaxies built as described in Section~\ref{sec:synthetic_galaxies} and Table~\ref{tab:PySynGal_params}. Each galaxy contains between $\Nhii=100$ and about 3000 constituent \hii regions, from which average \hii-region parameters are computed by weighting the individual \hii-region parameters by the intrinsic \Ha luminosities of these regions (see Section~\ref{sub:integproperties} for more details and comments on the distribution shapes).}
\label{fig:parameters_distribution_PySynGal}
\end{figure*}

We show in Fig.~\ref{fig:parameters_distribution_PySynGal} the distributions of integrated properties of the \Nsyngals composite star-forming galaxies built as described in the previous section. These include in particular the number of constituent \hii regions (\Nhii) and the integrated, dust-corrected \Ha luminosity (\LHagal), which incorporates the DIG contribution (Section~\ref{sec:sub:diffuseISM}). Also shown are average \hii-region parameters, computed by weighting the parameters of individual \hii regions by the intrinsic \Ha luminosities of these regions (\LHa; Section~\ref{sec:synthetic_galaxies}). In what follows, we refer to these average parameters as the `\Ha-weighted \hii-region parameters'.

The distributions of the various quantities in  Fig.~\ref{fig:parameters_distribution_PySynGal} can be understood from the way in which the parameters are sampled in Table~\ref{tab:PySynGal_params}. For example, the relation between the baseline \logOH and \logU (equation~\ref{eq:logU}), together with the upper limit on \LHa (equation~\ref{eq:PropaIncluded}) and the fact that the rate of H-ionizing photons, $Q_0$, scales as $U^3$ (Section~\ref{sec:sub:irregulargrids}), implies a deficit of galaxies with low O/H. Similarly, the geometric form factor \fr also depends on \logU through $Q_0$ and the limit on \LHa. The shapes of the \XO distributions in  Fig.~\ref{fig:parameters_distribution_PySynGal} originate from the dependence of \XHtot on \OHtot according to \citet{2017Nicholls_mnra466}. Most notably, the sharp increase in \COtot with \OHtot over the range $8.3\la\logOH\la9.0$ and the plateau values outside this range imply that for $\log(\OH)\la-3.7$ (corresponding to most of the range in Fig.~\ref{fig:parameters_distribution_PySynGal}), galaxies stack up at low \CO and are more sparsely sampled at higher \CO, corresponding to $-3.7\la\logOH\la3.0$. The trend is less marked for \NO, due to the adoption of a smooth dependence of \NOtot on \OHtot by \citet[][their equation~9]{2017Nicholls_mnra466}, while \SOtot, \NeOtot and \ArOtot are assumed to vary at most negligibly with \logOH (their table~2). Other parameters, such as \nh and A$_V$, have more symmetric distributions, sharply peaked near the middle of the fixed ranges over which individual \hii-region parameters are drawn for all galaxies. We note the uneven distribution of \Nhii, which arises from the random drawing of \LHa combined with the imposed \Ha luminosity function (equation~\ref{eq:PropaIncluded}).

\begin{figure*}
\includegraphics[width=18cm, trim=0 0 230 0, clip]{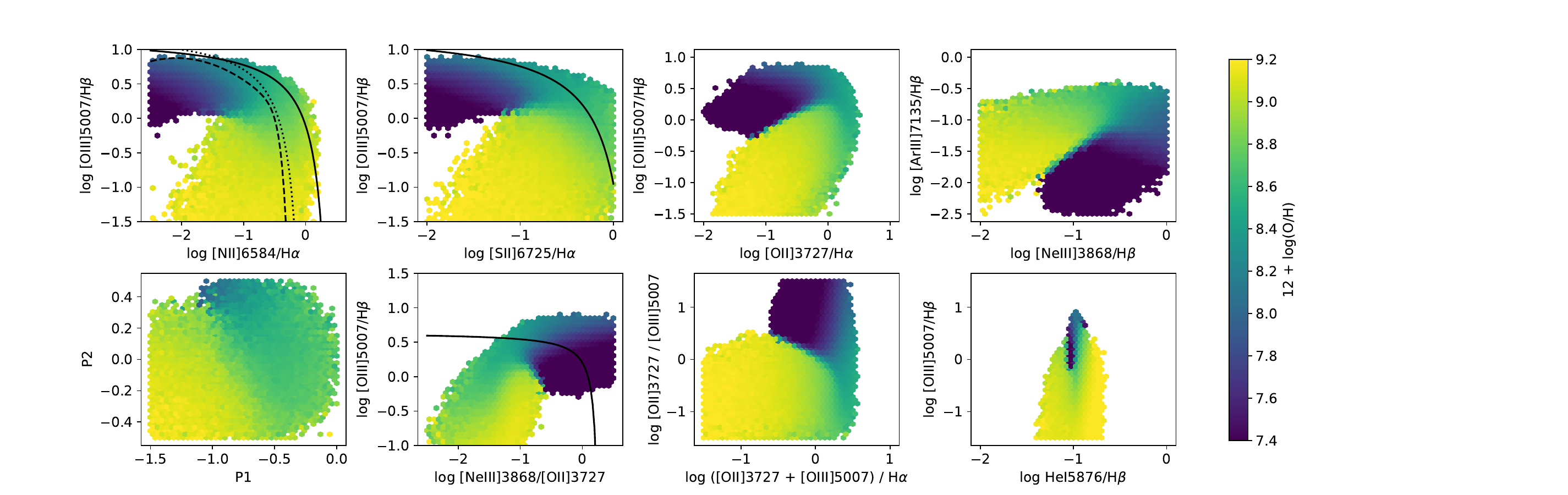}
\caption{Distribution of the \Nsyngals composite star-forming galaxies of Fig.~\ref{fig:parameters_distribution_PySynGal} in various line-diagnostic diagrams, colour-coded according to oxygen abundance as indicated on the right-hand side. In the \forba{O}{iii}{5007}/\Hb-versus-\forba{N}{ii}{6584}/\Ha (BPT) diagram, the separation criteria between AGNs and star-forming galaxies by \citet[solid line]{2001Kewley_apj556}, \citet[dotted line]{2003Kauffmann_mnra346} and \citet[dashed line]{2006Stasinska_mnra371} are shown. Similar criteria by \citet{2001Kewley_apj556} appear as solid lines in the \forba{O}{iii}{5007}/\Hb-versus-\forba{S}{ii}{6716,31}/\Ha and \forba{O}{iii}{5007}/\Hb-versus-\forba{O}{i}{6300}/\Ha (VO87) diagrams, and by \citet{2022Backhaus_apj926} in the \forba{O}{iii}{5007}/\Hb-versus-\forba{Ne}{iii}{3869}/\forba{O}{ii}{3727} \citep{2015Zeimann_apj798} diagram. See text Section~\ref{sub:integproperties} for details. }
\label{fig:BPTs_full}
\end{figure*}

It is of particular interest to check how the \Nsyngals composite star-forming galaxies built in this way populate standard line-diagnostic diagrams. This is shown in Fig.~\ref{fig:BPTs_full} for different diagrams involving \forba{O}{ii}{3727}, \forba{Ne}{iii}{3869}, \Hb, \forba{O}{iii}{5007}, \forba{O}{i}{6300}, \Ha, \forba{N}{ii}{6584}, \forba{S}{ii}{6725}, P1 and P2 (footnote~\ref{foot:plines}), inspired from various previous studies \citep[BPT, VO87,][]{2022Ji_aap659, 2015Zeimann_apj798}. In the \forba{O}{iii}{5007}/\Hb-versus-\forba{N}{ii}{6584}/\Ha (BPT) diagram, we show the criteria of \citet[solid line]{2001Kewley_apj556}, \citet[dotted line]{2003Kauffmann_mnra346} and \citet[dashed line]{2006Stasinska_mnra371} to separate AGN-dominated from star-forming galaxies. Solid lines show similar criteria by \citet{2001Kewley_apj556} in the \forba{O}{iii}{5007}/\Hb-versus-\forba{S}{ii}{6725}/\Ha and \forba{O}{iii}{5007}/\Hb-versus-\forba{O}{i}{6300}/\Ha (VO87) diagrams, and by \citet{2022Backhaus_apj926} in the \forba{O}{iii}{5007}/\Hb-versus-\forba{Ne}{iii}{3869}/\forba{O}{ii}{3727} \citep{2015Zeimann_apj798} diagram. 

The composite star-forming galaxies models fall roughly on the expected side of these criteria in all diagrams, except in the case of the \forba{O}{iii}{5007}/\Hb-versus-\forba{Ne}{iii}{3869}/\forba{O}{ii}{3727} diagram. This is not surprising, as the \citet{2022Backhaus_apj926} criterion in this diagram marks the region below which AGN-dominated galaxies are unlikely to be found, but does not exclude star-forming galaxies on either side (see their figure~6 and section~3.2; as well as \citealt{2023Trump_apj945, 2024Hu_apj971}). 

The models in Fig.~\ref{fig:BPTs_full}, which are colour-coded according to oxygen abundance, show well-defined sequences in \logOH in all diagrams, as expected from many previous studies \citep[e.g.][]{1979Pagel_mnras189, 1991McGaugh_apj380, 2001Charlot_mnra323, 2001Kewley_apj556}. Figure~\ref{fig:BPTs_full11} of Appendix~\ref{sec:app:BPT_full_1} shows analogues to Fig.~\ref{fig:BPTs_full} colour-coded according to other quantities: the DIG contribution to \Hb luminosity (DIGFrac), and the offsets $\varrho(\delta_\mathrm{N/O})$ and $\varrho(\delta_\mathrm{U})$ of the baseline galactic N/O and \logU with respect to canonical relations (Section~\ref{sec:sub:gal_param_distribution}). These show, for example, that star-forming galaxies located on the AGN side of the \citet{2001Kewley_apj556} relation in the BPT diagram tend to have large DIGFrac and $\varrho(\delta_\mathrm{N/O})$.

\subsection{O/H determination from integrated line ratios}
\label{sub:abundsGal}

We now examine how the oxygen abundance of a composite star-forming galaxy derived using the classical `direct method' \citep[e.g.][]{1959Aller_apj130, 1967Peimbert_apj150, 1969Peimbert_Bole5} differs from the true \Ha-weighted oxygen abundance of its constituent \hii regions.

We derive the electronic temperature and density of the emitting gas in a given galaxy using the \pyneb code \citep[version 1.1.19,][]{2015Luridiana_aap573, 2020Morisset_Atom8}, adopting the same atomic data as used to generate the \cloudy v17.03 photoionization models in Section~\ref{sub:testhii}, whose references are listed in Table~\ref{tab:atomic_data}.
\pyneb incorporates the effects of collisional excitation for lines emitted by metals. We use here a simplified two-zone ionization structure. The electronic temperatures of the N$^{+}$ and O$^{2+}$ zones are derived from the ratios of auroral to nebular forbidden lines \rforba{N}{ii}{5755}{6584} and \rforba{O}{iii}{4363}{5007}, and the electronic density of the S$^{+}$ zone from the ratio of forbidden lines \rforba{S}{ii}{6716}{6731}. \pyneb computes the O$^{+}$/H$^+$ and O$^{2+}$/H$^+$ ionic abundances assuming that the electronic density is the same everywhere and that the electronic temperature of the N$^{+}$ zone is representative of that of the O$^{+}$ zone. The sum of O$^{+}$/H$^+$ and O$^{2+}$/H$^+$ gives the total O/H abundance of the galaxy, assuming that the contributions of O$^0$ and O$^{3+}$ are negligible. No correction is applied to take into account emission process other than collisional excitation (e.g. recombination or fluorescence).

\begin{figure}
\includegraphics[width=8cm]{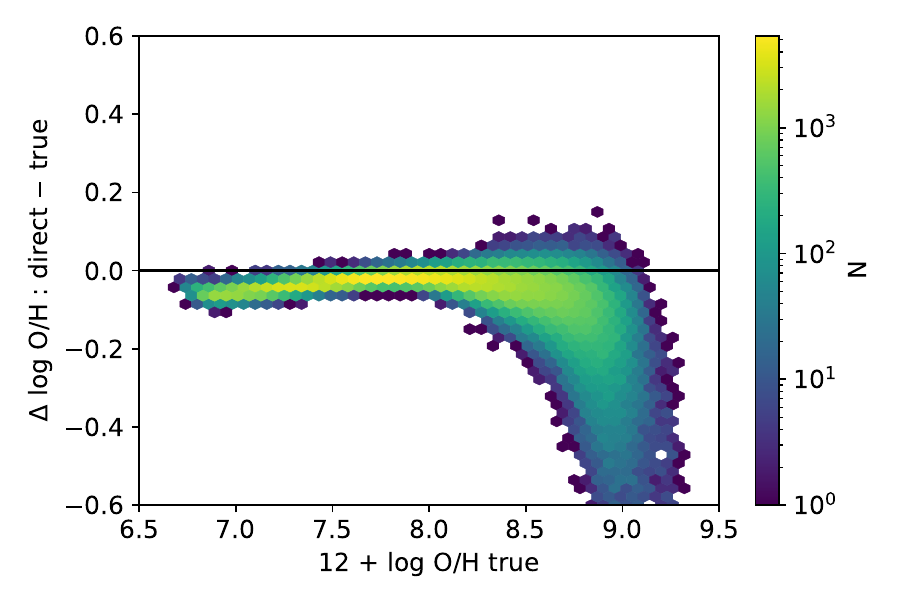}
\includegraphics[width=8.25cm]{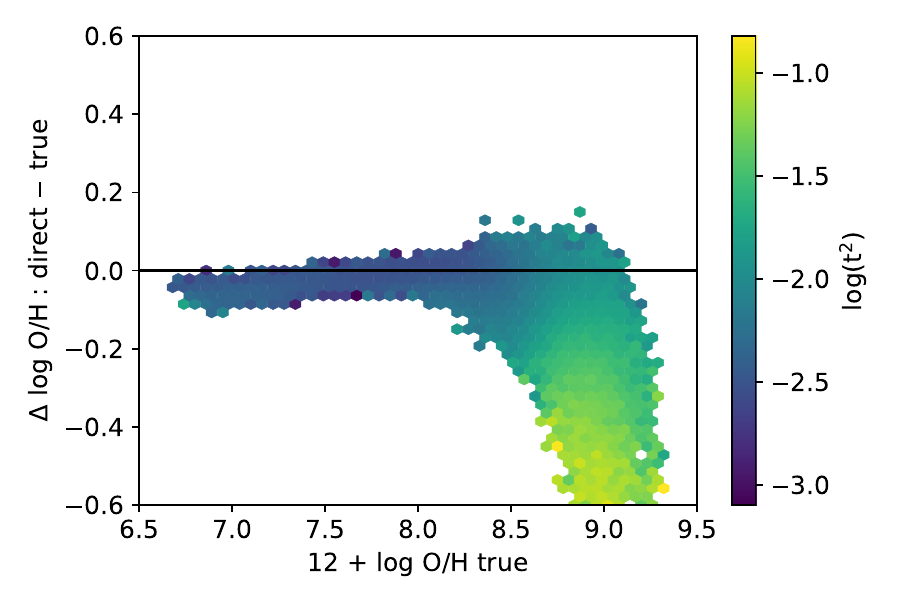}
\caption{Distribution of the logarithmic difference between \OH derived from the direct method and true, \Ha-weighted \OH as a function of \logOH, for the subset of \NbrightOiii composite star-forming galaxies of Fig.~\ref{fig:parameters_distribution_PySynGal} with $\log(\forba{O}{iii}{4363}/\Hb) > -3$. Top: colour-coded as a function of model number density. Bottom: colour-coded as a function of root-mean-square $t$ of \hii-region temperature fluctuations (equation~\ref{eq:Te_rms}).}

\label{fig:abundsHII}
\end{figure}

In the upper panel of Fig.~\ref{fig:abundsHII}, we show a density map of the logarithmic difference between O/H derived in this way and the true, \Ha-weighted O/H as a function of \logOH, for the subset of \NbrightOiii composite star-forming galaxies of Fig.~\ref{fig:parameters_distribution_PySynGal} with significant \forba{O}{iii}{4363} emission, i.e. with $\log(\forba{O}{iii}{4363}/\Hb) > -3$. We notice that at the lowest metallicities, i.e. for $\logOH\la7.5$, the direct method tends to slightly underestimate the oxygen abundance, by about 0.05\,dex. This result, reminiscent of the early finding by \citet{1999Kobulnicky_apj514}, is likely caused by the neglect of O$^{3+}$ in our \pyneb estimates, which starts to be significant as the ionization parameter increases with decreasing O/H (equation~\ref{eq:logU}).

Most notably, Figure~\ref{fig:abundsHII} shows that direct-method determinations of O/H in metal-rich galaxies can seriously underestimate the true O/H, by factors of up to several. 
Several origins for this phenomenon have been proposed, among them the neglect by the direct method, at high metallicity (i.e. low electronic temperature), of the contribution of recombination to the \forba{N}{ii}{5755} and \forba{O}{iii}{4363} auroral lines \citep[][and references therein]{2020Gomez-Llanos_mnra498} and the fact that any temperature gradient within the nebula preferentially enhances \forba{O}{iii}{5007} emission in low-temperature regions, while \forba{O}{iii}{4363} is more prominent in high-temperature regions \citep{2005Stasinska_aap434}. Moreover, our modelling of the electronic-temperature (\Te) distribution of \hii regions in composite galaxies offers a unique opportunity to study the influence of \Te fluctuations on O/H estimates of star-forming galaxies using the direct method (even though our \cloudy calculations do not include such fluctuations within individual regions). By analogy with \citet{1967Peimbert_apj150}'s parametrization of temperature fluctuations inside \hii regions, we define the root-mean-square $t$ of temperature fluctuations (weighted by \Ha luminosity) among \hii regions in a composite star-forming galaxy by
\begin{equation}
t^2 = \left\langle\left(\frac{\Te-T_0}{T_0}\right)^2\right\rangle_\mathrm{LH\alpha}\,,
\label{eq:Te_rms}
\end{equation}
where \LHa is the \Ha luminosity of the region of electronic temperature \Te, and 
\begin{equation}
T_0  = \langle\Te\rangle_\mathrm{LH\alpha}\,
\label{eq:Tzero}
\end{equation}
is the mean, \Ha-weighted temperature of \hii regions in the galaxy (Section~\ref{sub:integproperties}). The $t^2$ distribution of the galaxies in our sample is shown in the bottom-right panel of Fig.~\ref{fig:parameters_distribution_PySynGal}.

The lower panel of Fig.~\ref{fig:abundsHII} shows the analogue of the upper panel, but colour-coded according to $\log(t^2)$. Remarkably, this plot reveals that the galaxies for which the direct method most strongly underestimates O/H at high metallicities exhibit the strongest \Te fluctuations, with deficits of over 0.2\,dex corresponding to $t^2 > 0.03$. This result suggests that the spread in \Te among the constituent \hii regions of a galaxy might be an important factor causing the direct method to underestimate O/H. This echoes the conclusions drawn by \citet{2023Cameron_mnra522} from the radiation-hydrodynamics simulation of an isolated dwarf galaxy. 

Hence, Fig.~\ref{fig:abundsHII} illustrates the uncertainties and biases that can affect O/H determinations when obtained for observations in which light from numerous individual \hii regions and DIG are collected together. A detailed investigation of how these effects influence abundance determinations in samples of observed star-forming galaxies (and, a fortiori, abundances derived from stacking several galaxy spectra), as well as the impact of noise on these results, will be addressed in a forthcoming study.

\begin{table}
\caption{Atomic data sets used for collisionally-excited lines in the \cloudy and \pyneb calculations presented in this paper. \label{tab:atomic_data}}
\resizebox{\columnwidth}{!}{%
\begin{tabular}{lcc}
\hline
Ion & Transition Probabilities & Collision Strengths \\
\hline
N$^{+}$   &  \citet{2001Tachiev_Cana79} & \citet{2011Tayal_apjs195}\\
O$^{+}$   &  \citet{2004Froese-Fischer_Atom87} & \citet{2009Kisielius_mnra397}\\
O$^{2+}$  &  \citet{2001Tachiev_Cana79} & \citet{2014Storey_mnra441}\\
S$^{+}$   &  \citet{2014Kisielius_apj780} & \citet{2010Tayal_apjs188}\\
\hline
\end{tabular}
}
\end{table}

\section{Summary and perspectives}\label{sec:conclusion}

We have developed a novel approach to model the nebular emission from star-forming galaxies, incorporating the contributions of numerous individual \hii regions and a DIG component \citep[considering only DIG produced by HOLMES and excluding contributions from photon leakage from \hii regions, which may dominate in at least some star-forming galaxies; see][]{2024Lugo-Aranda_mnra528}. The approach leverages a machine-learning-based regression ANN, trained on a grid of models generated by the photoionization code \cloudy, to predict the emission-line properties of \hii regions under various physical conditions. 

The ANN can perform one million predictions of 14 line ratios, electron temperature and \Ha luminosity can be generated in less than 0.2~second from sampling the 13 input parameters, on a NVIDIA RTX A5000 GPU graphics card.
This \cloudy substitute can be applied in scenarios requiring the generation of numerous models, such as MCMC simulations and generic algorithms involving multiple model iterations.

We constructed \Nsyngals synthetic star-forming galaxies by summing, for each galaxy, the emission from between 100 and about 3000 \hii regions following an \Ha luminosity function of slope $\aLF=-1.9$ \citep{2018Rousseau-Nepton_mnra477, 2022Santoro_aap658}, each with different physical parameters and dust attenuation, and a component of diffuse gas photoionized by HOLMES \citep{2011Flores-Fajardo_mnra415}. The distributions of \hii-region parameters are tunable and chosen to reproduce observed trends, the relationships between \OH, \logU, \CO, \NO, \NeO, \SO and \ArO being considered as adjustable parameters changing from one galaxy to another. The integrated properties for each synthetic galaxy are defined as \Ha-weighted mean values of the corresponding properties of the contributing \hii regions. The integrated properties of the resulting \Nsyngals galaxies lie roughly in the regions expected to be populated by star-forming galaxies in standard line-ratio diagnostic diagrams involving \forba{O}{ii}{3727}, \forba{Ne}{iii}{3869}, \Hb, \forba{O}{iii}{5007}, \forba{O}{i}{6300}, \Ha, \forba{N}{ii}{6584} and \forba{S}{ii}{6725} \citep[BPT, VO87,][]{2022Ji_aap659, 2015Zeimann_apj798}. This approach enables one to explore how different parameter distributions affect the integrated emission-line properties of composite star-forming galaxies. 

As a practical application, we examined how the oxygen abundances of composite star-forming galaxies determined using the traditional ‘direct method’ \citep[e.g.][]{1959Aller_apj130, 1967Peimbert_apj150, 1969Peimbert_Bole5} compare to the actual \Ha-weighted abundances derived from the properties of the individual \hii regions that constitute galaxies. We find that the direct method exhibits systematic biases at both low ($\logOH<7.5$) and high ($\logOH>8.5$) metallicities, with a more pronounced effect at the higher end, resulting in underestimates of \OH in both cases. The bias at low metallicity  \citep[reminiscent of that found in the pioneer study of][]{1999Kobulnicky_apj514} arises from the common omission of O$^{3+}$ in \OH estimates achieved in the two-zone (O$^+$/O$^{2+}$) approximation using the standard \pyneb package \citep[][]{2015Luridiana_aap573, 2020Morisset_Atom8}. At high metallicities, the bias is most severe in galaxies with the largest spread in electronic temperatures among their constituent \hii\ regions, aligning with the findings of \citet{2023Cameron_mnra522} from radiation-hydrodynamics simulations of an isolated dwarf galaxy. These findings highlight the need to account for the composite nature of star-forming galaxies when deriving physical parameters from the observed integrated nebular emission.

The results presented in this paper are primarily illustrative, the main objective being to introduce our novel approach to modelling efficiently the integrated properties of composite star-forming galaxies. We have limited here our investigation to the optical emission-line properties of galaxies, but similar analyses could be extended to include ultraviolet and infrared emission-line diagnostics  \citep[e.g.][]{2019Cormier_aap626, 2019Hirschmann_mnra487}. This would involve training a new ANN to predict the relevant line luminosities and generating composite galaxy models by summing these additional line emissions among the constituent \hii regions.

In a forthcoming paper, we will apply the approach introduced here to fiting observed line ratios of star-forming galaxies. We plan to explore a new parametrization of the ANN model to avoid generating unrealistic combinations of input parameters which lead to improbable \Ha emissions. We also plan to use more efficient distributions for the input parameters, like the Latin Hypercube, the Sobol or the Halton ones. Additionally, we may update the training using models obtained from the \cloudy C23.01 version. 

\section*{Acknowledgements}
The authors would like to thank the anonymous referee for their valuable report, which significantly contributed to improving the quality of this paper.
The initial stages of this work were conceived and developed during a CM sabbatical at the Institut d'Astrophysique de Paris (IAP) to collaborate with SC. CM gratefully acknowledges the PASPA grant that supported this sabbatical. 
The authors thank Grazyna Stasi\'nska for valuable discussions. This work used v2.2.1 of the Binary Population and Spectral Synthesis (BPASS) models as described in \citet{2017Eldridge_pasa34} and \citet{2018Stanway_mnra479}. CM acknowledges the support of UNAM/DGAPA/PAPIIT grants IN101220 and IG101223.

\section*{Software}

In this study the following python libraries are used: numpy \citep{2020Harris_Natu585}, scipy \citep{2020Virtanen_Natu17}, matplotlib \citep{2007Hunter_Comp9}, astropy \citep{2013The-Astropy-Collaboration_ArXi, 2018Astropy-Collaboration_aj156}, PyNeb \citep{2015Luridiana_aap573, 2020Morisset_Atom8}, pyCloudy \citep{2014Morisset_}, sqlalchemy \citep{2012Bayer_}, pandas \citep{2010McKinney_, reback2020pandas}, scikit-learn \citep{2011Pedregosa_Jour12}, and Tensorflow \citep{2015MartinAbadi_}. This paper has been edited using Overleaf facilities.

\section*{Data Availability}


The Python scripts and notebooks used for this research are available from the github repository at \href{https://github.com/Morisset/pysyngal}{https://github.com/Morisset/pysyngal} (locked until publication of this and the forthcoming paper). The repository offers a downloadable version of the trained ANN, as well as the training set and training script. The original \cloudy models can be found on 3MdB\_17 under reference HII\_24. 



\bibliographystyle{mnras}
\bibliography{extracted} 



\appendix


\section{Using a neural network as a regressor for a grid of \cloudy \hii region models}
\label{sec:app:ANN}

The Machine Learning method used to substitute \cloudy in the prediction of emission-line ratios is of ANN type. This ANN is implemented using the Python TensorFlow library (\href{http://tensorflow.org/}{http://tensorflow.org/}). The neural architecture counts three hidden layers of 256, 512 and 256 activation cells, respectively. All activation functions are of GELU type \citep{2016Hendrycks_arXi}, which combines the fast-convergence benefits of ReLU and ELU functions with the regularization effect of a dropout, achieved by randomly setting some activations to zero during training. The GELU function is infinitely differentiable. The input layer accepts a 13-dimensional tensor, and the output layer generates a 16-dimensional tensor (see Section~\ref{sec:ML} for the physical parameters and observables associated with the input and output). The loss function is the negative logarithmic likelihood of the Gaussian distribution where the mean and standard deviation are estimated. This choice allows us to describe the data as noisy observations with heteroscedastic Gaussian noise. In the homoscedastic case, only the mean is estimated, and one recovers the classical MSE loss.

The optimizer is Adam \citep{2014Kingma_arXi} with parameters $\beta_1 = 0.9$ and $\beta_2 = 0.999$. A learning-rate scheduler is tailored using an alternation of exponential decays, starting at epoch 500. 
The goal is to maximize the rate at which the loss decreases while ensuring loss minimization with a learning rate ending at almost zero. Its value starts at 0.001 and ends at approximately $10^{-7}$.

Training is performed with a batch size of 2000, using 80 per cent of the grid of one-million \cloudy models described in Section~\ref{sec:grid}, with ionizing-cluster age $t_\mathrm{c}<7\,$Myr and $\HbFrac > 0.65$. We explore the impact of changing the training-set size, batch size and number of epochs. The results are presented in Table~\ref{tab:ANNs} for seven combinations of training-set sizes and number of epochs. The standard deviations listed in the table are obtained for test points in the triangle defined in Section~\ref{sec:sub:quality_regressor}.

Training times are provided for an NVIDIA RTX A5000 graphics card. Standard deviations of the difference between predicted and true values for each output tensor element are reported. The quality of predictions improves with larger training sets and more epochs. The ANN trained with 800,000 models over 8000 epochs achieves an impressive prediction capacity of close to 1 per cent in dex. This ANN is used for the analysis in the present paper. The input tensor is scaled using the scikit-learn StandardScaler function. The ANN is managed through the AI4Neb library. 

\begin{table*}
\caption{Performance of different ANNs with different training-set sizes and numbers of epochs. The training set consists of 80 per cent of the total data, while 20 per cent is used to evaluate the fit quality. The table lists the standard deviation (times 100) of the difference between predicted and true values (both being logarithm values of line ratios, \Te or \Ha) for each element of the prediction vector in the test set. The last row shows the mean of these 16 standard deviations.}
\label{tab:ANNs}
\centering
\begin{tabular}{lcccccccc}
\hline

Total set size [$\times10^5$]    & 2    & 4    & 6    & 6    & 8    & 10   & 10   & 10    \\
epochs [$\times10^3$]              & 2    & 2    & 2    & 10   & 2    & 2    & 8    & 8     \\
Batch size [$\times10^3$]         &  2 & 2 & 2 & 2 & 2 & 2 & 2 & 0.5 \\
Training time [mn]         &  5.8 & 11.6 & 14.5 & 72.1 & 19.4 & 23.5 & 88.3 & 345.0 \\
\hline
ANN output  & \multicolumn{8}{c}{standard deviation (x 100)}\\
\hline
\Ha/\Hb           & 0.2 & 0.2 & 0.2 & 0.1 & 0.2 & 0.2 & 0.1 & 0.1 \\
\forba{O}{iii}{5007}/H$\beta$        & 4.7 & 3.4 & 2.6 & 2.2 & 2.1 & 1.8 & 1.3 & 1.3 \\
\forba{N}{ii}{6584}/H$\beta$        & 4.4 & 3.2 & 2.5 & 2.2 & 2.2 & 1.9 & 1.6 & 1.5 \\
\forba{O}{ii}{3727}/H$\beta$        & 5.0 & 3.8 & 3.1 & 2.8 & 2.6 & 2.2 & 1.7 & 1.7 \\
\rforba{O}{iii}{4363}{5007}        & 3.6 & 2.9 & 2.3 & 2.1 & 2.0 & 1.8 & 1.3 & 1.3 \\
\rforba{N}{ii}{5755}{6584}        & 3.4 & 2.6 & 2.1 & 1.9 & 1.8 & 1.6 & 1.4 & 1.3 \\
\forba{S}{ii}{6725}/H$\beta$        & 4.4 & 3.1 & 2.5 & 2.3 & 2.2 & 1.9 & 1.7 & 1.5 \\
\forba{S}{iii}{9530}/H$\beta$        & 3.1 & 2.2 & 1.7 & 1.5 & 1.4 & 1.2 & 0.9 & 0.9 \\
\forba{Ar}{iii}{7135}/H$\beta$        & 3.6 & 2.6 & 2.0 & 1.7 & 1.7 & 1.4 & 1.1 & 1.0 \\
\forba{Ne}{iii}{3868}/H$\beta$        & 6.2 & 4.4 & 3.4 & 2.9 & 2.8 & 2.4 & 1.7 & 1.7 \\
\rforba{S}{ii}{6716}{6731}        & 0.5 & 0.4 & 0.4 & 0.2 & 0.3 & 0.3 & 0.2 & 0.2 \\
\alloa{He}{i}{5876}/H$\beta$        & 1.1 & 1.0 & 0.9 & 0.8 & 0.9 & 0.8 & 0.8 & 0.8 \\
\rforba{S}{iii}{6312}{9530}       & 3.4 & 2.5 & 2.0 & 1.7 & 1.6 & 1.4 & 1.1 & 1.1 \\
\forba{O}{ii}{7325}/H$\beta$        & 5.5 & 4.2 & 3.4 & 3.1 & 2.8 & 2.3 & 1.8 & 1.8 \\
\Te                       & 1.0 & 0.8 & 0.6 & 0.5 & 0.6 & 0.5 & 0.4 & 0.4 \\
L\Ha               & 2.1 & 1.9 & 1.4 & 0.9 & 1.4 & 1.4 & 0.8 & 0.8 \\
\hline
Mean                        & 3.3  & 2.5 & 1.9 & 1.7 & 1.7 & 1.5 & 1.1 & 1.1 \\
\hline

\end{tabular}
\end{table*}

\section{Impact of selected parameters on emission-line ratios of \hii-region models}
\label{sec:app:HIIregion}

\begin{figure*}
\includegraphics[width=17cm, trim={0 0 2.5cm 0},clip]{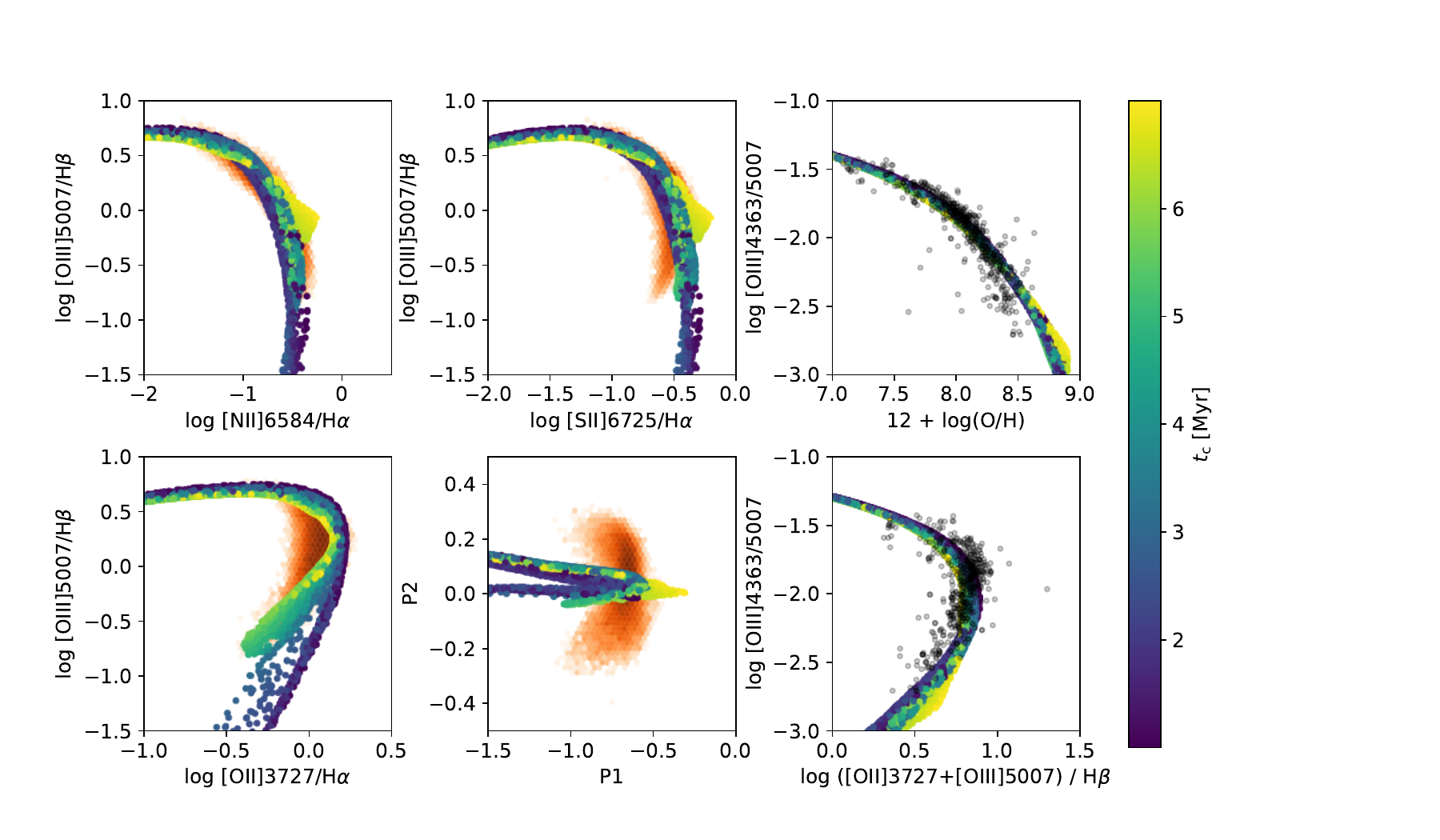}
\includegraphics[width=17cm, trim={0 0 2.5cm 0},clip]{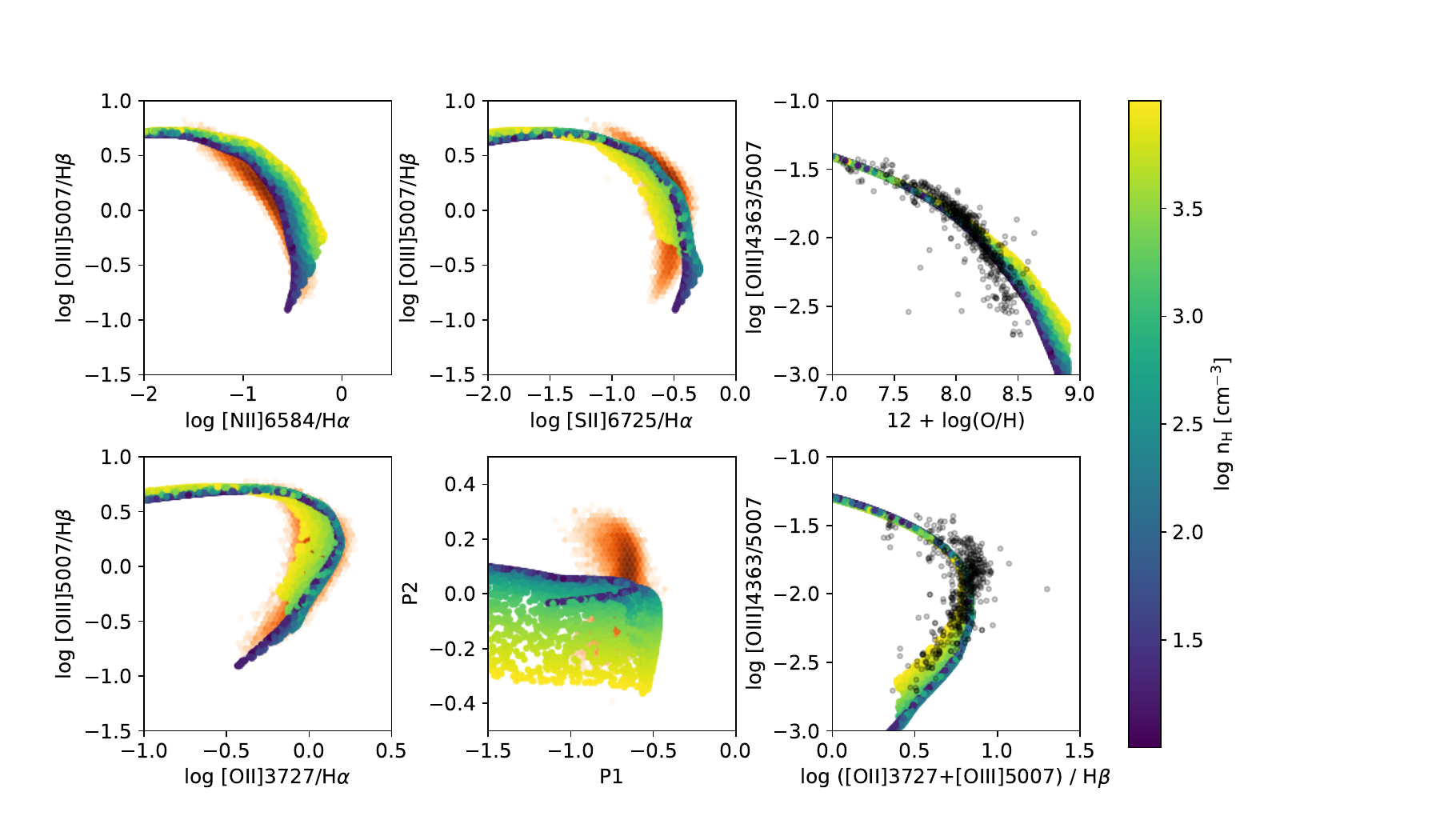}
\caption{Same as the six rightmost panels of Fig.~\ref{fig:Olaws}, but colour-coded according to ionizing-cluster age $t_\mathrm{c}$ (top) and gas density \nh (bottom).}
\label{fig:hii_preds_1}
\end{figure*}

\begin{figure*}
\includegraphics[width=17cm, trim={0 0 2.5cm 0},clip]{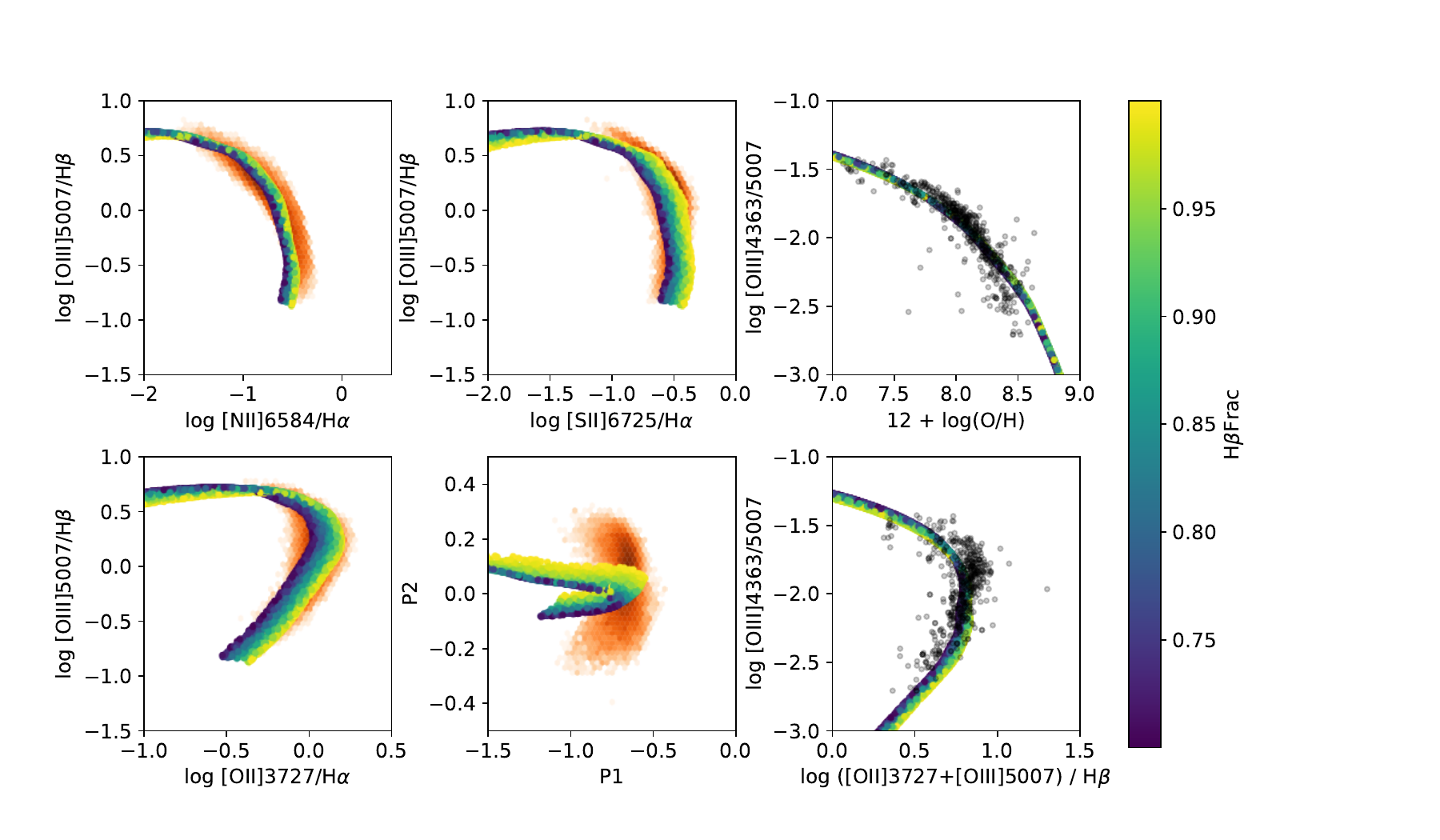}
\includegraphics[width=17cm, trim={0 0 2.5cm 0},clip]{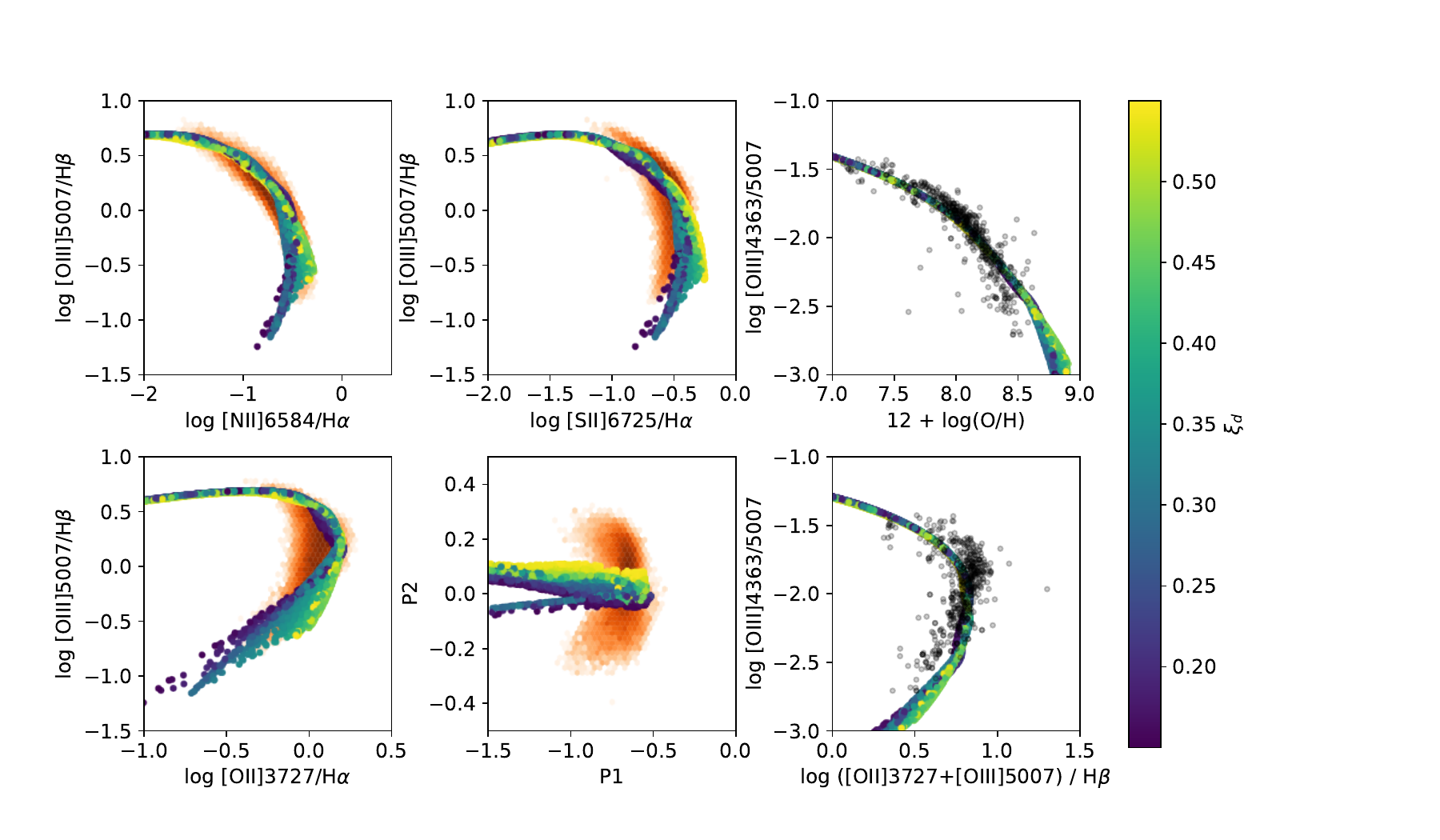}
\caption{Same as the six rightmost panels of Fig.~\ref{fig:Olaws}, but colour-coded according to \Hb-luminosity ratio \HbFrac of truncated versus radiation-bounded model (top) and dust-to-metal mass ratio \ksid (bottom).}
\label{fig:hii_preds_2}
\end{figure*}

Figures~\ref{fig:hii_preds_1} and \ref{fig:hii_preds_2} illustrate how variations in ionizing-cluster age, gas density and the \HbFrac and \ksid parameters (from top to bottom) impact the emission-line ratios in the same  diagrams as in the six rightmost panels of Fig.~\ref{fig:Olaws}. These figures reveal the complexity of these effects and the potential degeneracies that arise when attempting to reproduce the observed line ratios.

\section{Using a neural network as a regressor for a grid of \cloudy DIG models}
\label{sec:app:DIG-ANN}

The diffuse ionized gas is assumed to primarily consist of regions photoionized by HOLMES. To represent the DIG, we use the photoionization models of \citet{2023Martinez-Paredes_mnra525}, obtained from the 3MdB\_17 database under the reference `CB\_19'.
From this dataset, we select 9000 models with ages exceeding 1\,Myr and H densities $\nh\leq10\,\mathrm{cm}^{-3}$. We randomly split these models into a training set (80~per cent) and a test set (20~per cent). 

We train an artificial neural network, referred to as `DIG-ANN', using six input parameters: the age of the ionizing stars, drawn uniformly between 1 and 10\,Gyr; the O/H, C/O and N/O abundances in the DIG, assumed to match the global abundances of the composite galaxy (Section~\ref{sec:sub:gal_param_distribution}); the ionization parameter, drawn uniformly in the range $-4\leq\logU\leq-3$; and the quantity \HbFrac, drawn uniformly between 0.4 and 1.0. The outputs are the 14 emission-line ratios described in Section~\ref{sec:ML}. 

The DIG-ANN is implemented using the Scikit-Learn library \citep{2011Pedregosa_Jour12}, with three hidden layers comprising 50, 100 and 50 tanh-based perceptrons, respectively. Convergence is achieved using the LBFGS solver \citep{1989Liu_Math45}. The DIG-ANN is managed through the AI4Neb library (footnote~\ref{foot:AI4Neb}).

The standard deviations of the differences between the predicted and the test logarithmic values of the emission-line ratios are of the order of 0.02, comparable to the performance of the main ANN reported in Table~\ref{tab:ANNs}.


\section{Composite star-forming galaxies in emission-line diagrams colour-coded by different physical parameters}

\begin{figure*}
\includegraphics[width=18cm, trim=0 0 220 0, clip]{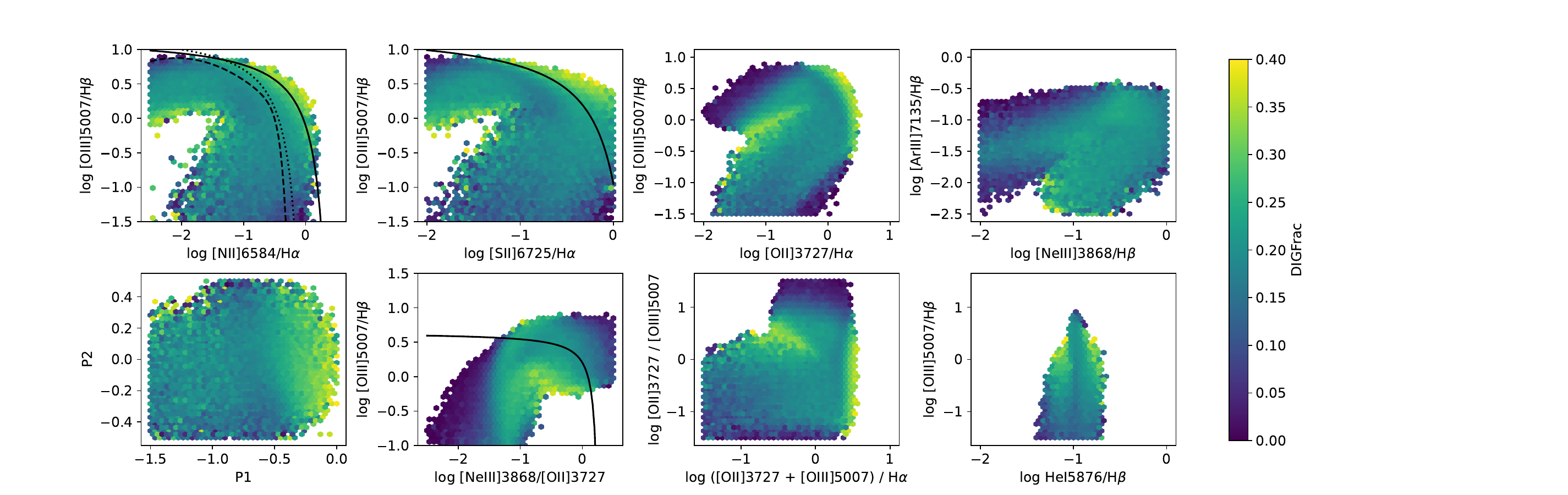}
\includegraphics[width=18cm, trim=0 0 220 0, clip]{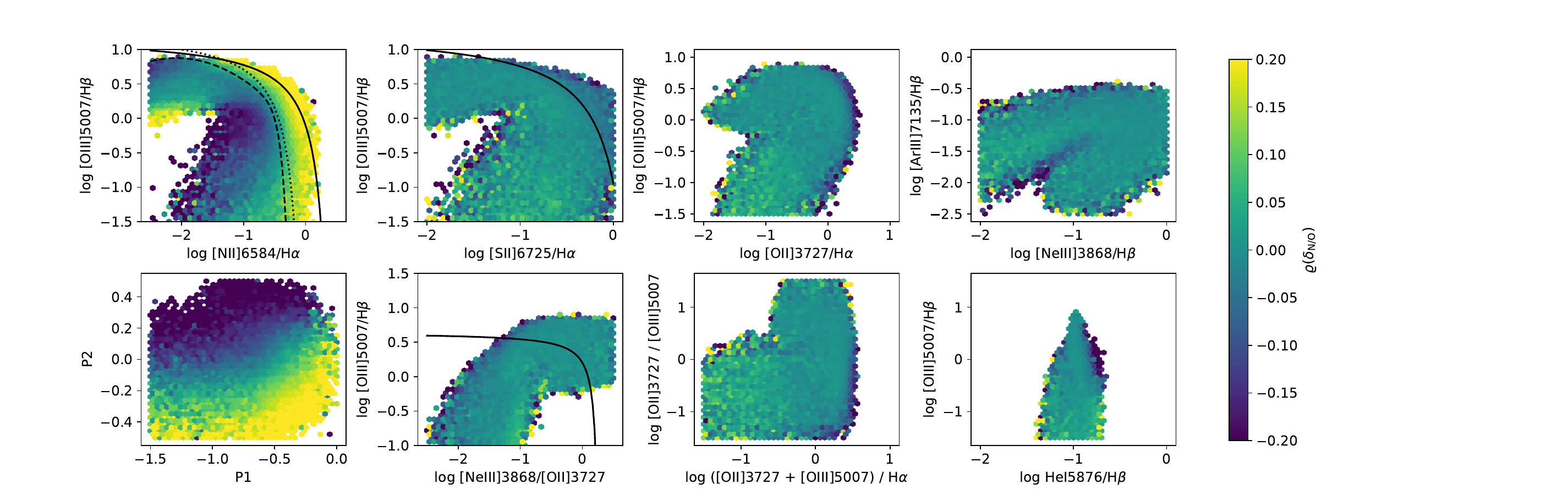}
\includegraphics[width=18cm, trim=0 0 220 0, clip]{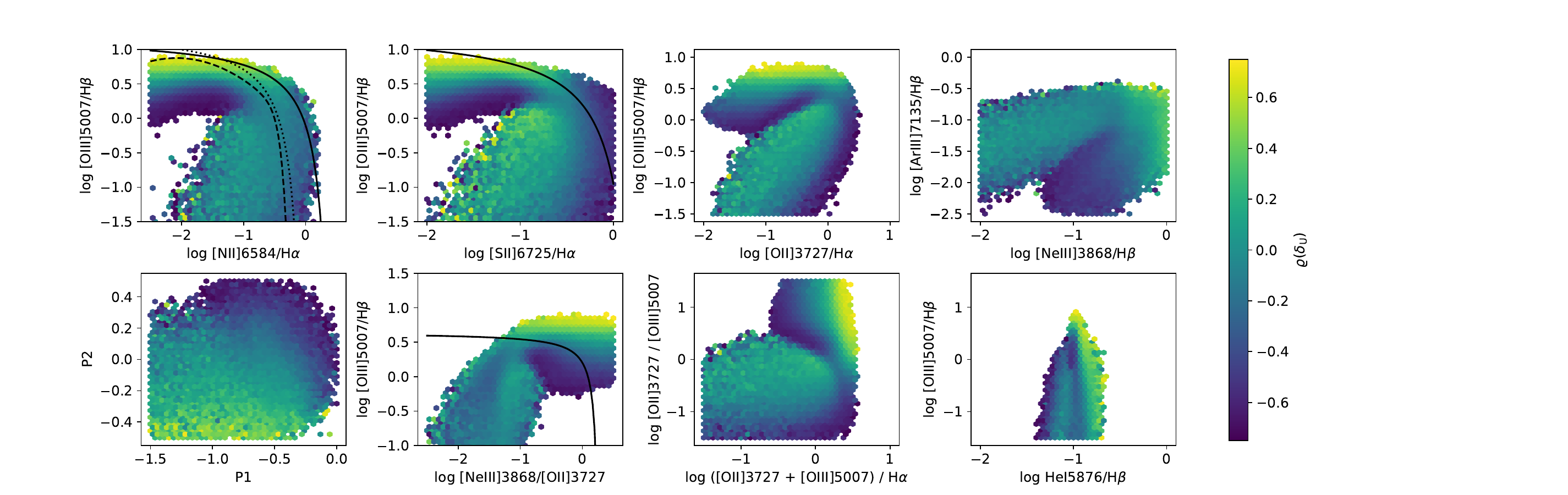}
\caption{Same as Fig.~\ref{fig:BPTs_full}, but colour-coded according to DIG contribution to \Hb luminosity (DIGFrac; top panels) and offsets $\varrho(\delta_\mathrm{N/O})$ (middle panels) and $\varrho(\delta_\mathrm{U})$ (bottom panels) of the baseline galactic \NO and \logU relative to the canonical relations described in Section~\ref{sec:sub:gal_param_distribution}. 
\label{fig:BPTs_full11}}
\end{figure*}

Figure~\ref{fig:BPTs_full11} shows analogues of Fig.~\ref{fig:BPTs_full} displaying the distribution of the \Nsyngals composite star-forming galaxies in various line-ratio diagnostic diagrams, but colour-coded according to the DIG contribution to \Hb luminosity (DIGFrac) and the offsets $\varrho(\delta_\mathrm{N/O})$ and $\varrho(\delta_\mathrm{U})$ of the baseline galactic \NO and \logU relative to canonical relations (Section~\ref{sec:sub:gal_param_distribution}). The results indicate, for example, that star-forming galaxies on the AGN side of the \citet{2001Kewley_apj556} relation in the \forba{O}{iii}{5007}/\Hb-versus-\forba{N}{ii}{6584}/\Ha (BPT) diagram generally exhibit high DIGFrac and $\varrho(\delta_\mathrm{N/O})$.

 
\label{sec:app:BPT_full_1}


\bsp	
\label{lastpage}
\end{document}